\documentclass[preprint,aps,tightenlines,showpacs,superscriptaddress]{revtex4}
\usepackage[dvips]{graphicx}
\usepackage{dcolumn}
\usepackage{bm}
\usepackage{epsfig}

\def\ohalf{{\textstyle{1\over 2}}}

\def\thalf{{\textstyle{3\over 2}}}

\newcommand{\beq}{\begin{equation}}
\newcommand{\eeq}{\end{equation}}

\begin{document}
\title{Dynamical Coupled-Channel Model of $\pi N$ Scattering \\
in the  $W \leq$ 2 GeV Nucleon Resonance Region\footnote{Notice: Authored by 
Jefferson Science Associates, LLC under U.S. DOE Contract No. DE-AC05-06OR23177. 
The U.S. Government retains a non-exclusive, paid-up, irrevocable, world-wide 
license to publish or reproduce this manuscript for U.S. Government purposes. 
} \\

\vspace{0.5cm}
(From EBAC, Thomas Jefferson National Accelerator Facility)} 
\author{B. Juli\'a-D\'{\i}az} 
\affiliation{ Excited Baryon Analysis Center (EBAC), Thomas Jefferson National
Accelerator Facility, Newport News, VA 22901, USA}
\affiliation{Departament d'Estructura i Constituents de la Mat\`{e}ria,
Universitat de Barcelona, E--08028 Barcelona, Spain}
\author{T.-S. H. Lee}
\affiliation{ Excited Baryon Analysis Center (EBAC), Thomas Jefferson National
Accelerator Facility, Newport News, VA 22901, USA}
\affiliation{Physics Division, Argonne National Laboratory, 
Argonne, IL 60439, USA}
\author{A. Matsuyama}
\affiliation{ Excited Baryon Analysis Center (EBAC), Thomas Jefferson National
Accelerator Facility, Newport News, VA 22901, USA}

\affiliation{Department of Physics, Shizuoka University, Shizuoka 422-8529, Japan}
\author{T. Sato}
\affiliation{ Excited Baryon Analysis Center (EBAC), Thomas Jefferson National
Accelerator Facility, Newport News, VA 22901, USA}
\affiliation{Department of Physics, Osaka University, Toyonaka, 
Osaka 560-0043, Japan}

\begin{abstract}
As a first step to analyze the electromagnetic meson production 
reactions in the nucleon resonance region, the parameters of 
the hadronic interactions of a dynamical coupled-channel model, 
developed in {\it Physics Reports 439, 193 (2007)}, are determined 
by fitting the $\pi N$ scattering data.
The channels included in the calculations 
are $\pi N$, $\eta N$ and $\pi\pi N$ which has $\pi\Delta$, $\rho N$, 
and $\sigma N$ resonant components. The non-resonant meson-baryon 
interactions of the model are derived from a set of Lagrangians 
by using a unitary transformation method. One or two bare excited nucleon
states in each of $S$, $P$, $D$, and $F$ partial waves are included 
to generate the resonant amplitudes in the fits. 
The parameters of the model are first determined by fitting as much as
 possible the empirical $\pi N$ elastic scattering amplitudes
of SAID up to 2 GeV. We then refine and confirm the resulting parameters by
directly comparing the predicted differential cross section and
target polarization asymmetry
with the original data of the elastic $\pi^{\pm} p \rightarrow \pi^{\pm} p$
and charge-exchange $\pi^- p \rightarrow \pi^0 n$ processes.
The predicted total cross sections of $\pi N$ reactions and
$\pi N\rightarrow \eta N$ reactions are also in good agreement 
with the data. 
Applications of the constructed 
model in analyzing the electromagnetic meson production data as well 
as the future developments are discussed.
\end{abstract}
\pacs{13.75.Gx, 13.60.Le,  14.20.Gk}
                                                                                
\maketitle

\section{Introduction}
It is now well recognized that a coupled-channel approach is needed to 
extract the nucleon resonance $(N^*)$ parameters from the data of $\pi N$ 
and electromagnetic meson production reactions. With the recent experimental 
developments~\cite{lee-reviewa,lee-reviewb}, such a theoretical effort is 
needed to analyze the very extensive data from Jefferson Laboratory (JLab), 
Mainz, Bonn, GRAAL, and Spring-8. To cope with this challenge, a dynamical 
coupled-channel model (MSL) for meson-baryon reactions in the 
nucleon resonance region has been developed recently~\cite{msl}. 
In this paper we report 
a first-stage determination of the parameters of 
this model by fitting the  $\pi N$ 
scattering data up to invariant mass $W$ = 2 GeV. 

The details of the MSL model are given in Ref.~\cite{msl}. Here we will only 
briefly recall its essential features. Similar to the earlier works on 
meson-exchange models~\cite{aay,aay-1,aa-1,aa-2,ap-87,afnan-88,pa-86,pj,gross,sl-1,sl-2,jlss07,
julich-1,julich-2,julich-3,julich-4,julich-5, ntuanl-1,ntuanl-2,pasc,afnan-1,afnan-2,fuda} of
pion-nucleon scattering, the starting point of the MSL model is a set of Lagrangians
describing the interactions between mesons ($M$ =$\gamma$, $\pi, \eta$ , $\rho, \omega$, 
$\sigma, \dots$) and baryons ($B = N, \Delta, N^*, \dots$). 
By applying a unitary transformation method~\cite{sl-1,sko}, an effective
Hamiltonian is then derived from the considered Lagrangian.
It can be cast into the following more transparent form
\begin{eqnarray}
H_{eff}= H_0 + \Gamma_V + v_{22}+ h_{\pi\pi N} \,,
\label{eq:H-1}
\end{eqnarray}
where
$H_0 = \sum_{\alpha}\sqrt{m_\alpha^2+\vec{p}_\alpha^{\,2}}$ with $m_\alpha$ denoting 
the mass of particle $\alpha$, and 
\begin{eqnarray}
\Gamma_V&=& \{ \sum_{N^*}
(\sum_{MB} \Gamma_{N^*\rightarrow MB} )
+ \sum_{M^*} h_{M^*\rightarrow \pi\pi} \} + \{c.c.\} \,,
\label{eq:gammav} \\
v_{22}&=& \sum_{MB,M^\prime B^\prime}v_{MB,M^\prime B^\prime}
+v_{\pi\pi} \,, 
\label{eq:v22} \\
h_{\pi\pi N} &=&  \sum_{N^*}\Gamma_{N^* \rightarrow \pi\pi N} 
+ \sum_{MB}[( v_{MB,\pi\pi N}) + (c.c.)]  +
v_{\pi\pi N, \pi\pi N}\,.
\label{eq:hpipin}
\end{eqnarray}
Here ${c.c.}$ denotes the complex conjugate of the terms on its
left-hand-side. In the above equations, $MB$ =  $\gamma N, \pi N$, 
$\eta N, \pi\Delta$, $\rho N, \sigma N$, represent the considered 
meson-baryon states. The resonance associated with the {\it bare} baryon 
state $N^*$ is induced by the vertex interactions $\Gamma_{N^* \rightarrow MB }$ 
and $\Gamma_{N^* \rightarrow \pi\pi N}$. Similarly, the {\it bare} meson states 
$M^* $ = $\rho$, $\sigma$ can develop into resonances through the vertex 
interaction $h _{M^*\rightarrow \pi\pi}$. Note that the masses $M^0_{N^*}$ 
and $m^0_{M^*}$ of the bare states $N^*$ and $M^*$ are the parameters 
of the model which must be determined by fitting the $\pi N$ and $\pi\pi$ 
scattering data. They differ from the empirically determined resonance 
positions by mass shifts which are due to the coupling of the bare 
states to the scattering states. The term $v_{22}$ contains 
the non-resonant meson-baryon interaction $v_{MB,M'B'}$ and $\pi\pi$ 
interaction $v_{\pi\pi}$. The non-resonant interactions involving $\pi\pi N$ 
states are in $h_{\pi\pi N}$. All of these interactions are {\it energy independent},
an important feature of the MSL formulation. 

We note here that the Hamiltonian defined above does not have a $\pi N \leftrightarrow N$ 
vertex. By applying the unitary transformation method, this un-physical process 
as well as any vertex interaction $A \leftrightarrow B + C$ with a mass relation
$m_A < m_B+m_C$ are eliminated from the considered Hilbert space and their effects 
are absorbed in the effective interactions $v_{22}$ and $h_{\pi\pi N}$.
This procedure defines the Hamiltonian in terms of physical nucleons and 
greatly simplifies the formulation of a unitary reaction model. 
In particular, the complications due to the nucleon mass and wavefunction 
renormalizations do not appear in the resulting scattering equations. This makes 
the numerical calculations involving the $\pi\pi N$ channel much more tractable 
in practice. The details of this approach are discussed in Refs.~\cite{sl-1,sko} 
as well as in the earlier works on $\pi NN$ interactions~\cite{lee-matsuyama}.

Starting from the above Hamiltonian, the coupled-channel equations for
$\pi N$ and $\gamma N$ reactions are then derived by using the standard 
projection operator technique~\cite{feshbach}, as given explicitly in 
Ref.~\cite{msl}. The obtained scattering equations satisfy the 
two-body ($\pi N, \eta N$, $\gamma N$) and three-body ($\pi\pi N$)
unitarity conditions. The $\pi\Delta$, $\rho N$ and $\sigma N$ resonant 
components of the $\pi \pi N$ continuum are generated dynamically by the 
vertex interaction $\Gamma_V$ of Eq.~(\ref{eq:gammav}). Accordingly, the 
$\pi\pi N$ cuts are treated more rigorously than the commonly used quasi-particle 
formulation within which these resonant channels are treated as simple 
two-particle states with a phenomenological parametrization of their widths.
The importance of such a dynamical treatment of unstable particle channels
was well known in earlier studies of $\pi N$ scattering~\cite{aay,thomas-review} 
and $\pi NN$ reactions~\cite{garcilazo}.

A complete determination of the parameters of the model Hamiltonian
defined by Eqs.(\ref{eq:H-1})-(\ref{eq:hpipin}) requires good fits
to all of the data of $\pi N$ and $\gamma N$ reactions up to invariant 
mass $W \leq $ about 2 GeV. Obviously, this is a very complex task 
and can only be accomplished step by step. Our strategy is as follows. 
We need to first determine the parameters associated with the 
hadronic interaction parts of the Hamiltonian. With the fits to 
$\pi\pi$ phase shifts in Ref.~\cite{johnstone}, the $\pi\pi$ 
interactions $h_{\rho,\pi\pi}$ and $h_{\sigma,\pi\pi}$ and the 
corresponding bare masses for $\rho$ and $\sigma$ have been determined 
in an isobar model with $v_{\pi\pi}=0$. We next proceed in two stages. 
The first-stage is to determine the ranges of the parameters of the
interactions $\Gamma_{N^*\rightarrow MB}$ and $v_{MB,M'B'}$.
This will be achieved by fitting the $\pi N$ scattering data from
performing coupled-channel calculations which neglect the more
 complex three-body interaction term
$h_{\pi\pi N}$. This simplification greatly reduces the numerical
complexity and the number of parameters to be determined in the fits.
This first-stage fit will provide the 
starting parameters to fit both the data of $\pi N$  scattering 
and $\pi N \rightarrow \pi\pi N$ reactions. In this second-stage, the 
parameters associated with $\Gamma_{N^*\rightarrow MB}$ and $v_{MB,M'B'}$ 
will be refined and the parameters of $h_{\pi\pi N}$ are then determined. 
The dynamical coupled-channel calculations for such more extensive fits are 
numerically more complex, as explained in Ref.~\cite{msl}.

In this work we report on the results from our first-stage determination 
of the parameters of $\Gamma_{N^*\rightarrow MB}$ and $v_{MB,M'B'}$ of 
Eqs.(\ref{eq:gammav})-(\ref{eq:v22}) 
with $MB, M'B' = \pi N, \eta N, \pi\Delta, \rho N,\sigma N$. 
We proceed in two steps. We first  locate the
range of the model parameters by fitting  as much as possible the
empirical $\pi N$ elastic scattering
amplitudes up to $W = 2$ GeV of SAID~\cite{said}.
We then refine and confirm the resulting parameters by
directly comparing our predictions with the original $\pi N$ scattering data.
Our procedures are similar to
what have been used in determining the nucleon-nucleon ($NN$)
potentials~\cite{v14} from
fitting $NN$ scattering data.

The constructed model can describe well almost all of
the empirical $\pi N$ amplitudes 
in $S$, $P$, $D$, and $F$ partial waves of SAID~\cite{said}.
We then show that the predicted differential cross sections and
target polarization asymmetry are in good agreement with
the original data of elastic $\pi^{\pm} p \rightarrow \pi^{\pm} p$ and
charge-exchange $\pi^- p \rightarrow \pi^0 n$ processes.
Furthermore 
the predicted total cross sections of the $\pi N$ reactions and 
$\pi N \rightarrow \eta N$ reactions agree well with the data. Thus the 
constructed model is at least comparable to, if not better than, all of 
the recent $\pi N$ models~\cite{pj,ntuanl-2,gross, sl-1, afnan-1,julich-4,
julich-5,pasc,fuda}. It can be used to perform a first-stage extraction 
of the $\gamma N \rightarrow N^*$ parameters by analyzing the photo- and 
electro-production of $single$ $\pi$ meson. It has also provided us with
a starting point for
performing the second-stage determination of the model parameters by
also fitting the data of $\pi N\rightarrow \pi\pi N$ reactions. 
Our efforts in these directions are in progress and will be reported elsewhere.

In Section II, we recall the coupled-channel equations presented in 
Ref.~\cite{msl}. The calculations performed in this work are described 
in Section III. The fitting procedure is described in Section IV 
and the results are presented in Section V.  In Section VI we give a 
summary and discuss future developments.

\section{Dynamical Coupled-channel equations}

\begin{figure}[t]
\centering
\includegraphics[width=12cm,angle=-0]{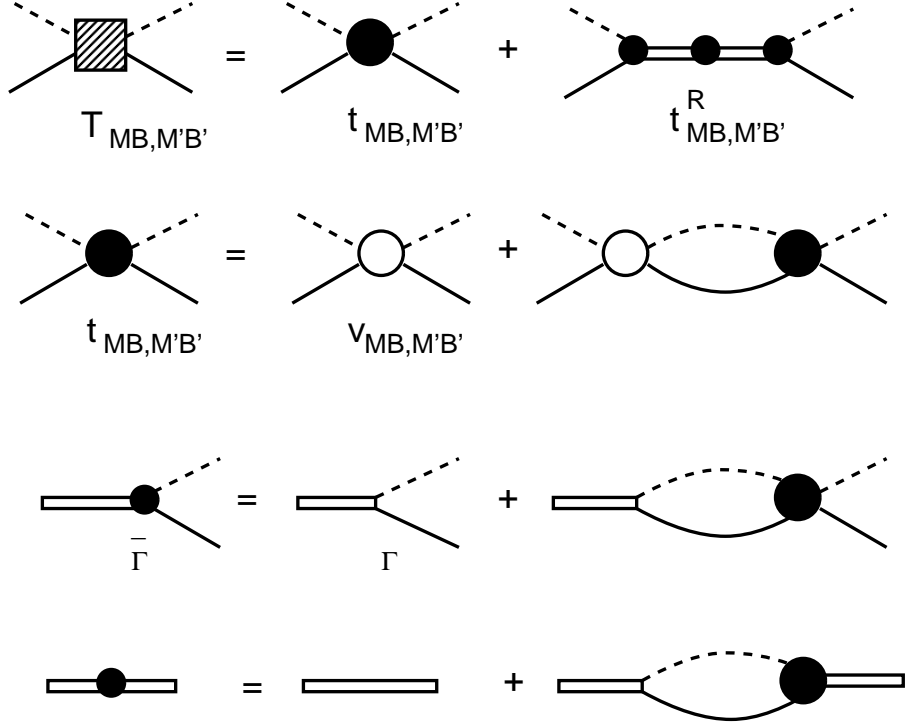}
\caption{Graphical representation of Eqs.(\ref{eq:tmbmb})-(\ref{eq:sig-sn}).}
\label{fig:tmatmbmb}
\end{figure}

\begin{figure}[t]
\includegraphics[width=12cm]{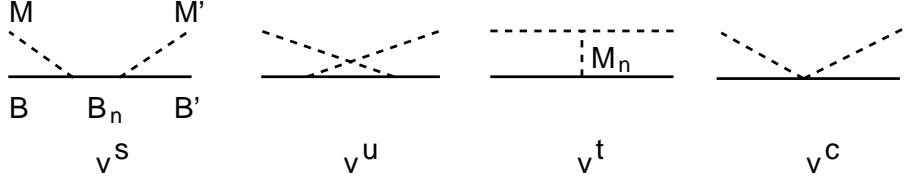}
\caption{Mechanisms for $v_{MB,M'B'}$ of Eq.~(\ref{eq:veff-mbmb}):
$v^s$ direct s-channel, 
$v^u$ crossed u-channel,
$v^t$ one-particle-exchange t-channel, 
$v^c$ contact interactions.
\label{fig:int}}
\end{figure}

\begin{figure}[t]
\centering
\includegraphics[width=12cm,angle=-0]{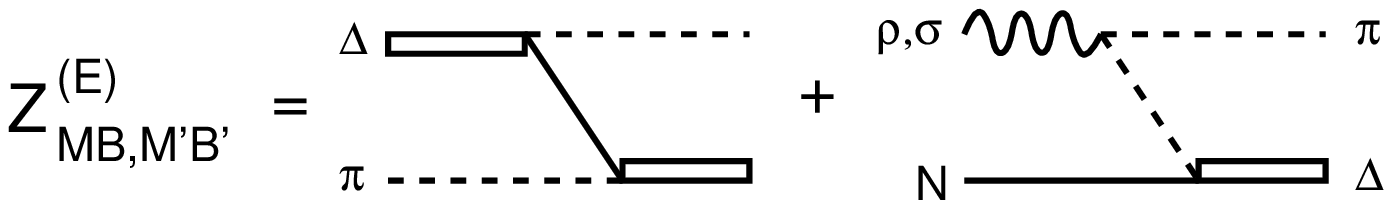}
\caption{One-particle-exchange interactions 
$Z^{(E)}_{\pi\Delta,\pi\Delta}(E)$, $Z^{(E)}_{\rho N,\pi\Delta}$
and $Z^{(E)}_{\sigma N,\pi\Delta}$ of Eq.~(\ref{eq:veff-mbmb}).}
\label{fig:z}
\end{figure}

With the simplification that $\pi\pi N$ interaction $h_{\pi\pi N}$ of
Eq.~(\ref{eq:hpipin}) is set to zero, the meson-baryon ($MB$) scattering 
equations derived in Ref.~\cite{msl} are illustrated in Fig.~\ref{fig:tmatmbmb}.
Explicitly, they are defined by the following equations
\begin{eqnarray}
 T_{MB,M^\prime B^\prime}(E)  &=&  
 t_{MB,M^\prime B^\prime}(E)
+ 
 t^R_{MB,M^\prime B^\prime}(E) \,,
\label{eq:tmbmb}
\end{eqnarray}
where $MB = \pi N, \eta N, \pi\Delta, \rho N, \sigma N$. The full  
amplitudes $T_{\pi N,\pi N}(E)$ can be directly used to calculate $\pi N$ 
scattering observables. The non-resonant amplitude $t_{MB,M^\prime B^\prime}(E)$ 
in Eq.~(\ref{eq:tmbmb}) is defined by the coupled-channel equations,
\begin{eqnarray}
t_{MB,M^\prime B^\prime}(E)= V_{MB,M^\prime B^\prime}(E)
+\sum_{M^{\prime\prime}B^{\prime\prime}}
V_{MB,M^{\prime\prime}B^{\prime\prime}}(E) \;
G_{M^{\prime\prime}B^{\prime\prime}}(E)    \;
t_{M^{\prime\prime}B^{\prime\prime},M^\prime B^\prime}(E)  \,
\label{eq:nr-tmbmb}
\end{eqnarray}
with
\begin{eqnarray}
V_{MB,M^\prime B^\prime}(E)= v_{MB,M^\prime B^\prime}
+Z^{(E)}_{{M}{B},{M}^\prime {B}^\prime}(E)\,. 
\label{eq:veff-mbmb}
\end{eqnarray}
Here the interactions $v_{MB,M'B'}$ are derived from the tree-diagrams 
illustrated in Fig.~\ref{fig:int} by using a unitary transformation
method~\cite{sl-1,sko}. It is energy independent and free of singularity. 
On the other hand, $Z^{(E)}_{{M}{B},{M}^\prime {B}^\prime}(E)$ 
is induced by the decays of the unstable particles ($\Delta$, $\rho$, 
$\sigma$) and thus contains  {\it moving} singularities due to the
$\pi\pi N$ cuts, as illustrated in Fig.\ref{fig:z}. Here we note that 
if the $\pi\pi N$ interaction term $h_{\pi\pi N}$ of Eq.(\ref{eq:hpipin}) 
is included, the driving term Eq.~(\ref{eq:veff-mbmb}) will have an 
additional term $Z^{(I)}_{{M}{B},{M}^\prime {B}^\prime}(E)$
which involves a 3-3 $\pi\pi N$ amplitude 
$t_{\pi\pi N,\pi\pi N}$, as given in Ref.~\cite{msl},
and hence is much more difficult to calculate.
As explained in Section I, we neglect this term in this first-stage
fit to the $\pi N$ scattering data.

The second term in the right-hand-side of Eq.~(\ref{eq:tmbmb}) 
is the resonant term defined by
\begin{eqnarray} 
t^R_{MB,M^\prime B^\prime}(E)= \sum_{N^*_i, N^*_j}
\bar{\Gamma}_{MB \rightarrow N^*_i}(E) [D(E)]_{i,j}
\bar{\Gamma}_{N^*_j \rightarrow M^\prime B^\prime}(E) \,,
\label{eq:tmbmb-r} 
\end{eqnarray}
with
\begin{eqnarray}
[D^{-1}(E)]_{i,j} = (E - M^0_{N^*_i})\delta_{i,j} - \bar{\Sigma}_{i,j}(E)\,,
\label{eq:nstar-g}
\end{eqnarray}
where $M_{N^*}^0$ is the bare mass of the resonant state $N^*$, and the 
self-energies are 
\begin{eqnarray}
\bar{\Sigma}_{i,j}(E)= \sum_{MB}\Gamma_{N^*_i\rightarrow MB} G_{MB}(E)
\bar{\Gamma}_{MB \rightarrow N^*_j}(E) \,.
\label{eq:nstar-sigma}
\end{eqnarray}
The dressed vertex interactions in Eq.~(\ref{eq:tmbmb-r}) and
Eq.~(\ref{eq:nstar-sigma}) are (defining 
$\Gamma_{MB\rightarrow N^*}=\Gamma^\dagger_{N^* \rightarrow MB}$)
\begin{eqnarray}
\bar{\Gamma}_{MB \rightarrow N^*}(E)  &=&  
{ \Gamma_{MB \rightarrow N^*}} + \sum_{M^\prime B^\prime}
t_{MB,M^\prime B^\prime}(E) 
G_{M^\prime B^\prime}(E)
\Gamma_{M^\prime B^\prime \rightarrow N^*}\,, 
\label{eq:mb-nstar} \\
\bar{\Gamma}_{N^* \rightarrow MB}(E)
 &=&  \Gamma_{N^* \rightarrow MB} +
\sum_{M^\prime B^\prime} \Gamma_{N^*\rightarrow M^\prime B^\prime}
G_{M^\prime B^\prime }(E)t_{M^\prime B^\prime,M B}(E) \,. 
\label{eq:nstar-mb}
\end{eqnarray}
It is useful to mention here that if there is only one $N^*$
in the considered partial wave, the resonant amplitude 
(Eq.~(\ref{eq:tmbmb-r})) can be written as
\begin{eqnarray}
t^R_{MB,M^\prime B^\prime}(E)= \frac{\bar{\Gamma}_{MB \rightarrow N^*_1}(E)
\bar{\Gamma}_{N^*_1 \rightarrow M^\prime B^\prime}(E)}
{E- E_R(E) +i \frac{\Gamma_R(E)}{2}}
\label{eq:bw-form}
\end{eqnarray}
with
\begin{eqnarray}
E_R(E)&=& M^0_{N^*} + {\rm Re} [\bar{\Sigma}(E)] \,, \label{eq:bw-er} \\
\Gamma_R(E) &=& - 2 \, {\rm Im} [\bar{\Sigma}(E)] \label{eq:bw-sd} \,,
\end{eqnarray}
where,
\begin{eqnarray}
\bar{\Sigma}(E) 
= \sum_{MB}\Gamma_{N^*\rightarrow MB} G_{MB}(E)
\{\sum_{M'B'}[\delta_{MB,M'B'} +
t_{MB,M'B'}(E) G_{M'B'}(E)]
\}{\Gamma}_{M'B' \rightarrow N^*}(E) \,. \nonumber \\
\label{eq:bw-gamma}
\end{eqnarray}
The form Eq.~(\ref{eq:bw-form}) is similar to the commonly used
Breit-Wigner form, but the resonance position $E_R(E)$ and width $\Gamma_R(E)$
are determined by the $N^* \rightarrow MB$ vertex and the non-resonant 
amplitude $t_{MB,M'B'}$. This is the consequence of the unitarity 
condition and is an important and well known feature of a dynamical 
approach. Namely, the resonance amplitude necessarily includes the 
non-resonant mechanisms. This feature is consistent with the well 
developed formal reaction theory~\cite{feshbach}. Eq.~(\ref{eq:bw-gamma}) 
indicates that it is essential to understand the non-resonant mechanisms
in extracting the bare vertex functions $\Gamma_{N^*,MB}$ which
contain the information for exploring the $N^*$ structure. The 
parameterization used for $\Gamma_{N^*,MB}$ will be
explained in Section~\ref{sec:cal}. We also note here that the 
energy dependence of $E_R(E)$ and $\Gamma_R(E)$, defined by 
Eqs~(\ref{eq:bw-er})-(\ref{eq:bw-sd}), is essential in determining 
the resonance poles in the complex E-plane.

The meson-baryon propagators $G_{MB}$ in the above equations 
are 
\begin{eqnarray}
G_{MB}(k,E) = \frac{1}{E-E_M(k)-E_B(k)+ i\epsilon}
\end{eqnarray}
for the stable particle channels $MB=\pi N, \eta N$, and
\begin{eqnarray}
G_{MB}(k,E) = \frac{1}{E-E_M(k)-E_B(k)-\Sigma_{MB}(k,E)} 
\label{eq:rgreen}
\end{eqnarray}
for the unstable particle channels $MB=\pi\Delta, \rho N, \sigma N$.
The self-energies~\cite{propag} in Eq.~(\ref{eq:rgreen}) are
\begin{eqnarray}
\Sigma_{\pi\Delta}(k,E)&=& \frac{m_\Delta}{E_\Delta(k)}
\int q^2 dq
\frac{M_{\pi N}(q)}{[M^2_{\pi N}(q)+k^2]^{1/2}}
\frac{|f_{\Delta,\pi N}(q)|^2}
{E-E_\pi(k) -[(E_N(q)+E_\pi(q))^2+k^2]^{1/2} + i\epsilon} \,,  
\label{eq:sig-pid} \nonumber \\
& & \\
\Sigma_{\rho N}(k,E)&=&
\frac{m_\rho}{E_\rho(k)}\int q^2 dq
\frac{M_{\pi \pi}(q)}{[M^2_{\pi \pi}(q)+k^2]^{1/2}}
\frac{|f_{\rho,\pi\pi}(q)|^2}
{E-E_N(k) - [(2E_\pi(q))^2+k^2]^{1/2} + i\epsilon} \,, 
\label{eq:sig-rn}  \\
& & \nonumber \\
\Sigma_{\sigma N}(k,E)&=&
\frac{m_\sigma}{E_\sigma(k)}
\int q^2 dq
\frac{M_{\pi \pi}(q)}{[M^2_{\pi \pi}(q)+k^2]^{1/2}}
\frac{|f_{\sigma,\pi\pi}(q)|^2}
{E-E_N(k) -[(2 E_\pi (q))^2+k^2]^{1/2} + i\epsilon} \,,
\label{eq:sig-sn}
\end{eqnarray}
where $M_{\pi N}(q)=E_\pi(q)+E_N(q)$ and $M_{\pi\pi}(q)=2E_\pi(q)$. The 
vertex function $f_{\Delta,\pi N}(q)$ is taken from Ref.~\cite{sl-1},
$f_{\rho,\pi \pi}(q)$ and $f_{\sigma,\pi \pi}(q)$ are from the 
isobar fits~\cite{johnstone} to the $\pi\pi $ phase shifts. They are 
also given explicitly in~\cite{msl}.

Here we note that the driving term $Z^{(E)}_{MB,M'B'}$ of Eq.~(\ref{eq:veff-mbmb}) 
is also determined by the same vertex functions $f_{\Delta,\pi N}(q)$, 
$f_{\rho,\pi\pi}(q)$ and $f_{\rho,\pi\pi}(q)$ of 
Eqs.~(\ref{eq:sig-pid})-(\ref{eq:sig-sn}). This consistency is essential 
for the solutions of Eq.~(\ref{eq:nr-tmbmb}) to satisfy the unitarity condition.

\section{Calculations}
\label{sec:cal}

We solve the coupled-channel equations defined by 
Eqs.(\ref{eq:tmbmb})-(\ref{eq:sig-sn}) in the partial-wave 
representation. The input of these equations are the partial-wave 
matrix elements of $\Gamma_{N^*\rightarrow MB}$ and $v_{MB,M'B'}$ 
of Eqs.(\ref{eq:gammav})-(\ref{eq:v22}), with $MB,M'B'=\pi N, \eta N$, 
$\pi \Delta, \rho N, \sigma N$, and $Z^{(E)}_{MB,M'B'}$ of Eq.~(\ref{eq:veff-mbmb}) 
with $MB, M'B'=\pi \Delta, \rho N, \sigma N$. The calculations of these 
matrix elements have been given explicitly in the appendices of Ref.~\cite{msl}. 
Here we only mention a few points which are needed for later discussions.

In deriving the non-resonant interactions $v_{MB,M'B'}$ of Eq.~(\ref{eq:veff-mbmb})
we consider the tree-diagrams (Fig.~\ref{fig:int}) generated from
a set of Lagrangians with $\pi$, $\eta$, $\sigma$, $\rho$, $\omega$, 
$N$, and $\Delta$ fields. The higher mass mesons, such as $a_0$, $a_1$ 
included in other meson-exchange $\pi N$ models, such as
the J\"ulich model~\cite{julich-4}, are not considered. 
The employed Lagrangians are ( in the convention of Bjorken and Drell~\cite{bj})
\begin{eqnarray}
L_{\pi NN} &=& -\frac{f_{\pi NN}}{m_\pi}\bar{\psi}_N\gamma_\mu \gamma_5
\vec{\tau}\psi_N\cdot \partial^\mu\vec{\phi_\pi} \,, \label{eq:L-pinn} \\
& & \nonumber \\
L_{\pi N\Delta} &=& -\frac{f_{\pi N\Delta}}{m_\pi}\bar{\psi}_\Delta^\mu
\vec{T}\psi_N\cdot \partial_\mu\vec{\phi_\pi} \label{eq:L-pind}\,, \\
& & \nonumber \\
L_{\pi\Delta\Delta}&=&\frac{f_{\pi\Delta\Delta}}{m_\pi}\bar{\psi}_{\Delta\mu}
\gamma^\nu \gamma_5 \vec{T}_\Delta \psi_\Delta^\mu
\cdot \partial_\nu\vec{\phi}_\pi \,, \\
& &  \nonumber \\
L_{\eta NN} &=& -\frac{f_{\eta NN}}{m_\eta}\bar{\psi}_N\gamma_\mu \gamma_5
\psi_N \partial^\mu\phi_\eta \,. \\
& & \nonumber \\
L_{\rho NN} &=& g_{\rho NN}\bar{\psi}_N[\gamma_\mu -
\frac{\kappa_\rho}{2m_N}\sigma_{\mu\nu}\partial^\nu]
\vec{\rho^\mu} \cdot \frac{\vec{\tau}}{2}\psi_N \label{eq:L-rnn} \,, \\
& & \nonumber \\
L_{\rho N\Delta} &=& -i\frac{f_{\rho N\Delta}}{m_\rho}\bar{\psi}_\Delta^\mu
\gamma^\nu \gamma_5 \vec{T} \cdot
[\partial_\mu \vec{\rho_\nu}-\partial_\nu\vec{\rho_\mu}]\psi_N
+[h.c.] \,, \\
& & \nonumber \\
L_{\rho\Delta\Delta}&=&g_{\rho \Delta\Delta}\bar{\psi}_{\Delta\alpha}
[\gamma^\mu- \frac{\kappa_{\rho\Delta\Delta}}{2m_\Delta}
\sigma^{\mu\nu}\partial_\nu]
\vec{\rho_\mu}\cdot \vec{T}_\Delta \psi_\Delta^\alpha \,, \\
& & \nonumber \\
L_{\rho\pi\pi}&=& g_{\rho\pi\pi}[\vec{\phi_\pi}\times
\partial_\mu \vec{\phi_\pi}] \cdot \vec{\rho}^\mu\,, \\
& & \nonumber \\
L_{NN\rho\pi} &=& \frac{f_{\pi NN}}{m_\pi}g_{\rho NN}
\bar{\psi}_N\gamma_\mu \gamma_5
\vec{\tau}\psi_N\cdot \vec{\rho^\mu} \times \vec{\phi_\pi}
\label{eq:rho-ct1} \,, \\
& & \nonumber \\
L_{NN\rho\rho}&=&-\frac{\kappa_\rho g_{\rho NN}^2}{8m_N}
\bar{\psi}_N\sigma^{\mu\nu}\vec{\tau}
\psi_N \cdot \vec{\rho_\mu}\times \vec{\rho_\nu} \,. \label{eq:rho-ct2} \\
& & \nonumber \\
L_{\omega NN} &=& g_{\omega NN}\bar{\psi}_N[\gamma_\mu -
\frac{\kappa_\omega}{2m_N}\sigma_{\mu\nu}\partial^\nu]
{\omega^\mu}  \psi_N  \,, \\
& & \nonumber \\
L_{\omega\pi\rho}&=&-\frac{g_{\omega\pi\rho}}{m_\omega}
\epsilon_{\mu\alpha\lambda\nu}\partial^\alpha\vec{\rho^\mu}\partial^\lambda
\vec{\phi_\pi}\omega^\nu \,, \\
& & \nonumber \\
L_{\sigma NN} &=& g_{\sigma NN}\bar{\psi}_N \psi_N\phi_\sigma \\
L_{\sigma\pi\pi}&=& -\frac{g_{\sigma\pi\pi}}{2m_\pi}
 \partial^\mu \vec{\phi}_\pi\partial_\mu \vec{\phi}_\pi \phi_\sigma \,.
\label{eq:L-snn}
\end{eqnarray}

To solve the coupled-channel equations, Eq.~(\ref{eq:nr-tmbmb}),
we need to regularize the matrix elements of $v_{MB,M'B'}$, 
illustrated in Fig.~\ref{fig:int}. Here we follow Ref.~\cite{sl-1} 
in order to use the parameters determined in the $\Delta$ (1232) 
region as the starting parameters in our fits. For the $v^s$ and $v^u$ 
terms of Fig.~\ref{fig:int}, we include at each meson-baryon-baryon 
vertex a form factor of the following form
\beq
F(\vec{k},\Lambda)=[\vec{k}^2/[(\vec{k}^2+\Lambda^2)]^2
\label{eq:ff}
\eeq
with $\vec{k}$ being the meson momentum. For the meson-meson-meson 
vertex of $v^t$ of Fig.~\ref{fig:int}, the form Eq.~(\ref{eq:ff}) is 
also used with $\vec{k}$ being the momentum of the exchanged meson. 
For the contact term $v^c$, we regularize it by 
$F(\vec{k},\Lambda)F(\vec{k'},\Lambda')$.

With the non-resonant amplitudes generated from solving Eq.~(\ref{eq:nr-tmbmb}), 
the resonant amplitude $t^R_{MB,M'B'}$ Eq.~(\ref{eq:tmbmb-r}) then depends 
on the bare mass $M^0_{N^*}$ and the bare $N^*\rightarrow MB$ vertex functions.
As discussed in Ref.~\cite{msl}, these bare $N^*$ parameters can perhaps 
be taken from a hadron structure calculation which {\it does not} include 
coupling with meson-baryon continuum states or meson-exchange quark 
interactions. Unfortunately, such information is not available to us.
We thus use the following parameterization
\begin{eqnarray}
{\Gamma}_{N^*,MB(LS)}(k)
&=& \frac{1}{(2\pi)^{3/2}}\frac{1}{\sqrt{m_N}}C_{N^*,MB(LS)}
\left[\frac{\Lambda_{N^*,MB(LS)}^2}{\Lambda_{N^*,MB(LS)}^2
 + (k- k_R)^2}\right]^{(2+L/2)}
\left[\frac{k}{m_\pi}\right]^{L} \,.
\label{eq:gmb}
\end{eqnarray}
where $L$ and $S$ are the orbital angular momentum and the total spin
of the $MB$ system, respectively. The above parameterization
accounts for the threshold $k^L$ dependence and the right power $(2+L/2)$ 
such that the integration for calculating the dressed vertex 
Eq.~(\ref{eq:mb-nstar})-(\ref{eq:nstar-mb}) is finite. Nevertheless 
as we will discuss in Section~\ref{sec:res} this parameterization 
could be too naive.

The partial-wave quantum numbers for the considered channels are 
listed in Table~\ref{tab:pw}. The numerical methods for handling 
the moving singularities due to the $\pi\pi N$ cuts in $Z^{(E)}_{MB,M'B'}$ 
(Fig.~\ref{fig:z}) in solving Eq.~(\ref{eq:nr-tmbmb}) are explained
in detail in Ref~\cite{msl}. To get the $\pi N$ elastic scattering amplitudes,
we can use either the method of contour rotation by solving the equations
on the complex momentum axis $k=ke^{-i\theta}$ with $\theta >0$ or
the Spline-function method developed in Refs.~\cite{AM-1,AM-2} and 
explained in detail in Ref.~\cite{msl}. We perform the calculations using 
these two very different methods and they agree within less than $1\%$. When 
$Z^{(E)}_{MB,M'B'}$ is neglected, Eq.~(\ref{eq:nr-tmbmb}) can be solved 
by the standard subtraction method since the resonant propagators, 
Eqs.~(\ref{eq:rgreen}), for unstable particle channels $\pi\Delta$, 
$\rho N$, and $\sigma N$ are free of singularity on the real momentum 
axis. A code for this simplified case has also been developed to confirm 
the results from using the other two methods. 

The method of contour rotation becomes difficult at high $W$ since the
required rotation angle $\theta$ is very small. The Spline function method has
no such limitation and we can perform calculations at $ W > 1.9$
GeV without any difficulty. 
Typically, 24 and 32 mesh points are needed to get convergent solutions 
of the coupled-channel integral equation~(\ref{eq:nr-tmbmb}). Such mesh 
points are also needed to get stable integrations in evaluating the dressed 
resonance quantities Eqs.~(\ref{eq:nstar-sigma})-(\ref{eq:nstar-mb}).

\begin{table}
\begin{ruledtabular}
\begin{tabular}{c|c|c|c|c|c}
          &\multicolumn{5}{c}{$(LS)$ of the considered partial waves}\\
\hline
          & $\pi N$     & $\eta N$    & $\pi \Delta $   & $\sigma N$   & $\rho N$  \\
\hline
$S_{11}$  & ($0,\ohalf$) &($0,\ohalf$) &($2,\thalf$) &($1,\ohalf$) &($0,\ohalf$), ($2,\thalf$)\\
$S_{31}$  & ($0,\ohalf$) &   $-$       &($2,\thalf$) &    $-$
&($0,\ohalf$), ($2,\thalf$)\\
\hline
\hline
$P_{11}$  & ($1,\ohalf$) &($1,\ohalf$) &($1,\thalf$) &($0,\ohalf$) &($1,\ohalf$), ($1,\thalf$)\\
$P_{13}$  & ($1,\ohalf$) &($1,\ohalf$) &($1,\thalf$),($3,\thalf$)  &($2,\ohalf$) &($1,\ohalf$),($1,\thalf$), ($3,\thalf$)\\
$P_{31}$  & ($1,\ohalf$) &   $-$       &($1,\thalf$) &    $-$      &($1,\ohalf$), ($1,\thalf$)\\
$P_{33}$  & ($1,\ohalf$) &   $-$       &($1,\thalf$),($3,\thalf$)  &     $-$&($1,\ohalf$),($1,\thalf$), ($3,\thalf$)\\
\hline
\hline
$D_{13}$  & ($2,\ohalf$) &($2,\ohalf$) &($0,\thalf$),($2,\thalf$) &($1,\ohalf$)
&($2,\ohalf$), ($0,\thalf$), ($4,\thalf$)\\
$D_{15}$  & ($2,\ohalf$) &($2,\ohalf$) &($2,\thalf$) , ($4,\thalf$)&($3,\ohalf$) &($2,\ohalf$), ($2,\thalf$), ($4,\thalf$)\\
$D_{33}$  & ($2,\ohalf$) &    $-$      &($0,\thalf$),($2,\thalf$) & $-$
&($2,\ohalf$), ($0,\thalf$), ($2,\thalf$)\\
$D_{35}$  & ($2,\ohalf$) &    $-$      &($2,\thalf$), ($4,\thalf$) &   $-$ &($2,\ohalf$), ($2,\thalf$), ($4,\thalf$)\\
\hline
\hline
$F_{15}$  & ($3,\ohalf$) &($3,\ohalf$) &($1,\thalf$),($3,\thalf$) &($2,\ohalf$)
&($3,\ohalf$), ($1,\thalf$), ($3,\thalf$)\\
$F_{17}$  &     ($3,\ohalf$)& ($3,\ohalf$) & ($3,\thalf$),($5,\thalf$) 
& ($4,\ohalf$) & ($3,\ohalf$), ($3,\thalf$), ($5,\ohalf$)      \\
$F_{35}$  & ($3,\ohalf$) &    $-$      &($1,\thalf$),($3,\thalf$) &$-$         &($3,\ohalf$), ($1,\thalf$), ($3,\thalf$)\\
$F_{37}$  &   ($3,\ohalf$) &  $-$    & ($3,\thalf$),($5,\thalf$) & $$-$ $                   & ($3,\ohalf$), ($3,\thalf$), ($5,\thalf$)     \\
\end{tabular}
\caption{The orbital angular momentum $(L)$ and total spin ($S$)of
the partial waves included in solving the coupled channel
 Equation (\ref{eq:nr-tmbmb}). \label{tab:pw}}
\end{ruledtabular}
\end{table}

\section{Fitting Procedure}

With the specifications given in Section III, 
the parameters associated with $Z^{(E)}_{MB,M'B'}$ of Eq.~(\ref{eq:veff-mbmb})
are completely determined from fitting the $\pi\pi$ phase shifts in
Refs.~\cite{sl-1} and~\cite{johnstone}. Thus the considered 
model has the following parameters: 
(a) the coupling constants associated with the Lagrangians 
listed in Eqs.~(\ref{eq:L-pinn})-(\ref{eq:L-snn}),
(b) the cutoff $\Lambda$ for each vertex of $v_{MB,M'B'}$ (Fig.~\ref{fig:int}), 
(c) the coupling strength $C_{N^*,MB(LS)}$ and range $k_R$ 
and $\Lambda_{N^*,MB(LS)}$ of the bare $N^* \rightarrow MB$ vertex Eq.~(\ref{eq:gmb}), and 
(d) the bare mass $M^0_{N^*}$ of each $N^*$ state. We determine these by 
fitting the $\pi N$ scattering data.

Our fitting procedure is as follows. We first perform fits to the
$\pi N$ scattering data up to about 
1.4 GeV and including only one bare state, the $\Delta$ (1232) resonance.
In these fits, the starting coupling constant parameters of $v_{MB,M'B'}$ 
are taken from the previous studies of $\pi N$ and $NN$ scattering, which 
are also given in Ref.~\cite{msl}. Except the $\pi NN$ coupling constant 
$f_{\pi NN}$ all coupling constants and the cutoff parameters are allowed 
to vary in the $\chi^2$-fit to the $\pi N$ data.
The coupled-channel effects can shift the coupling constants greatly from 
their starting values. We try to minimize these shifts by allowing the 
cutoff parameters to vary in a very wide range 
500 MeV $ < \Lambda < $ 2000 MeV. 
Some signs of coupling constants, which could not be fixed by the previous 
works~\cite{nohl}, are also allowed to change. We then use the parameters 
from these fits at low energies as the starting ones to fit the amplitudes 
up to 2 GeV by also adjusting the resonance parameters, $M^0_{N^*}$,
$C_{N^*,MB(LS)}$, $k_R$ and $\Lambda_{N^*,MB(LS)}$. Here we need to specify 
the number of bare $N^*$ states in each partial wave. The simplest approach 
is to assume that each of 3-star and 4-star resonances listed by the Particle 
Data Group~\cite{PDG} is generated from a bare $N^*$ state of the model 
Hamiltonian Eq.~(\ref{eq:H-1}). However, this choice is perhaps not well 
justified since the situation of the higher mass $N^*$'s is not so clear.

We thus start the fits including only the bare states which generate the 
$lowest$ and well-established $N^*$ resonance in each partial wave. The 
second higher mass bare state is then included when a good fit can not 
be achieved. We also impose the condition that if the resulting $M^0_{N^*}$ 
is too high $  > 2.5 $ GeV, we remove such a bare state in the fit.
This is due to the consideration that the interactions due to such a heavy 
bare $N^*$ state could be just the separable representation of some non-resonant 
mechanisms which should be included in $v_{MB,M'B'}$. In some partial 
waves the quality of the fits is not very sensitive to the $N^*$ couplings 
to $\pi\Delta$, $\rho N$, and $\sigma N$. But the freedom of varying these 
coupling parameters is needed to achieve good fits.

It is rather difficult to fit all partial waves simultaneously because
the number of resonance parameters to be determined is very large.
We proceed as follows. We first fit only 3 or 4 partial waves which have
well established resonant states, and whose amplitudes have an involved energy
dependence. These are the $S_{11}$, $P_{11}$, $S_{31}$ and $P_{33}$
partial waves. These fits are aimed at identifying the possible ranges of
the parameters associated with $v_{MB,M'B'}$. This step is most difficult
and time consuming. We then gradually extend the fits to include more
partial waves. For some cases, the fits can be reached easily by simply
adjusting the bare $N^*$ parameters. But it often requires some adjustments
of the non-resonance parameters to obtain new fits. This procedure  has to
be repeated many times to explore the parameter space as much as we can.
We carry out this very involved numerical task by using the fitting code
MINUIT and the parallel computation facilities at NERSC in US and the
Barcelona Supercomputing Center in Spain.

The most uncertain part of the fitting is to handle the large number of
parameters associated with the bare $N^*$ states. Here the use of the
empirical partial-wave amplitudes from SAID is an essential step in the
fit.  It allows us to locate
the ranges of the $N^*$ parameters partial-wave by partial-wave for a given
set of the parameters for the non-resonant $v_{MB,M'B,}$. Even with
this, the information is far from complete for pinning down the
$N^*$ parameters.
Perhaps the $N^*$ parameters associated with the $\pi N$ state are reasonably
well determined in this fit to the $\pi N$ scattering data.
The parameters associated with $\eta N$, $\pi \Delta$, $\rho N$ and $\sigma N$
can only be better determined by also fitting to
the data of $\pi N \rightarrow \eta N$
and $\pi N \rightarrow \pi\pi N$ reactions. This
will be pursued in our second-stage calculations, as discussed in section I.

It is useful to note here that the leading-order effect due to $Z^{(E)}$ of 
the meson-baryon interaction Eq.~(\ref{eq:veff-mbmb}) on $\pi N$ elastic scattering is
\begin{eqnarray}
\delta v_{\pi N,\pi N} = \sum_{MB, M'B'=\pi\Delta,\rho N,\sigma N}
v_{\pi N, MB}G_{MB}(E)Z^{(E)}_{MB,M'B'} G_{M'B'}(E) v_{M'B',\pi N}\,.
\label{eq:vcorr}
\end{eqnarray}
We have found by explicit numerical calculations that $\delta v_{\pi N,\pi N}$ 
is much weaker than $v_{\pi N,\pi N}$ and hence the coupled channel effects due 
to $Z^{(E)}_{MB,M'B'}$ on $\pi N$ elastic scattering amplitude are weak.
One example obtained from our model 
is shown in Table \ref{tab:unitarity-1}. Thus we first perform 
the fits without including $Z^{(E)}$ term to speed up the computation. We 
then refine the parameters by including this term in the fits. 

\begin{table}[h]
\begin{ruledtabular}
\begin{tabular}{c|c|c|c|c|c}
\hline
 & Re$[t_{\pi N,\pi N}]$&  Re$[t_{\pi N,\pi N}(Z^{(E)}=0$)]&&  
Im$[t_{\pi N,\pi N}]$&  Im$[t_{\pi N,\pi N}(Z^{(E)}=0$)] \\
\hline
$S_{11}$ & $-$0.00481 &  $-$0.00557 & & 0.0841 & 0.0827 \\
$P_{11}$ &  0.0937    &       0.103 & & 0.636  &  0.640 \\
$P_{13}$ &  0.169     & 0.181       & & 0.275  &  0.275 \\
$D_{13}$  & 0.202     & 0.194       & & 0.299  &  0.309 \\
$D_{15}$ &  0.117     & 0.116       & & 0.0179 &  0.0179 \\
$F_{15}$ &  0.290     & 0.291       & &  0.157 &  0.155 \\
$F_{17}$ &  0.0360    & 0.0359      & & 0.00293&0.00289 \\
$S_{31}$ & $-$0.433   &$-$0.437     & &0.496   &0.504 \\
$P_{31}$  &$-$0.253   &$-$0.230     & &0.434   &0.448 \\
$P_{33}$ &  0.0506    & 0.0306      & &0.510   &0.457 \\
$D_{33}$ & $-$0.00504 & $-$0.0135   & & 0.106  &0.104 \\
$D_{35}$ &  0.0551    & 0.0551      & &0.0540  &0.0537 \\
$F_{35}$ &  $-$0.0214  & $-$0.0229   & &0.0259  &0.0283 \\
$F_{37}$ &  0.0625    &    0.0626   & &0.00502 &0.00512 \\
\hline
\end{tabular}
\caption{ The effect of $Z^{(E)}_{MB,M'B'}$ on the $\pi N$ scattering amplitudes
$t_{\pi N,\pi N}$ from solving Eq.~(\ref{eq:nr-tmbmb})
at $W = 1.7 $ GeV. The normalization is 
$t_{\pi N,\pi N} = (e^{2i\delta_{\pi N}}-1)/(2i)$, where 
$\delta_{\pi N}$ is the $\pi N$ scattering phase shift which could be complex
at energies above the $\pi$ production threshold.
\label{tab:unitarity-1}}
\end{ruledtabular}
\end{table}

\section{Results}
\label{sec:res}

As mentioned in section I,
we first locate the range of the parameters by
fitting the empirical $\pi N$ scattering amplitude of SAID~\cite{said}.
We then check and refine the resulting parameters
by directly comparing our predictions with the original $\pi N$ scattering
data.

\begin{figure}[tbp]
\vspace{20pt}
\begin{center}
\mbox{\epsfig{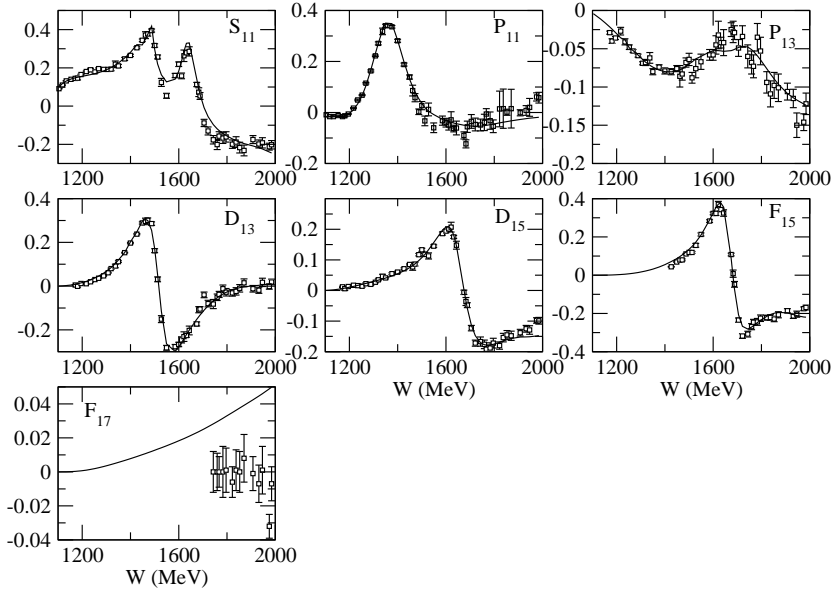}}
\end{center}
\caption{Real parts of the calculated $\pi N$ partial wave amplitudes 
(Eq.~(\ref{eq:tmbmb})) of isospin $T=1/2$  are compared with
the energy independent solutions of Ref.~\cite{said}.\label{figrealiso1}}
\end{figure}

\begin{figure}[tbp]
\vspace{20pt}
\begin{center}
\mbox{\epsfig{file=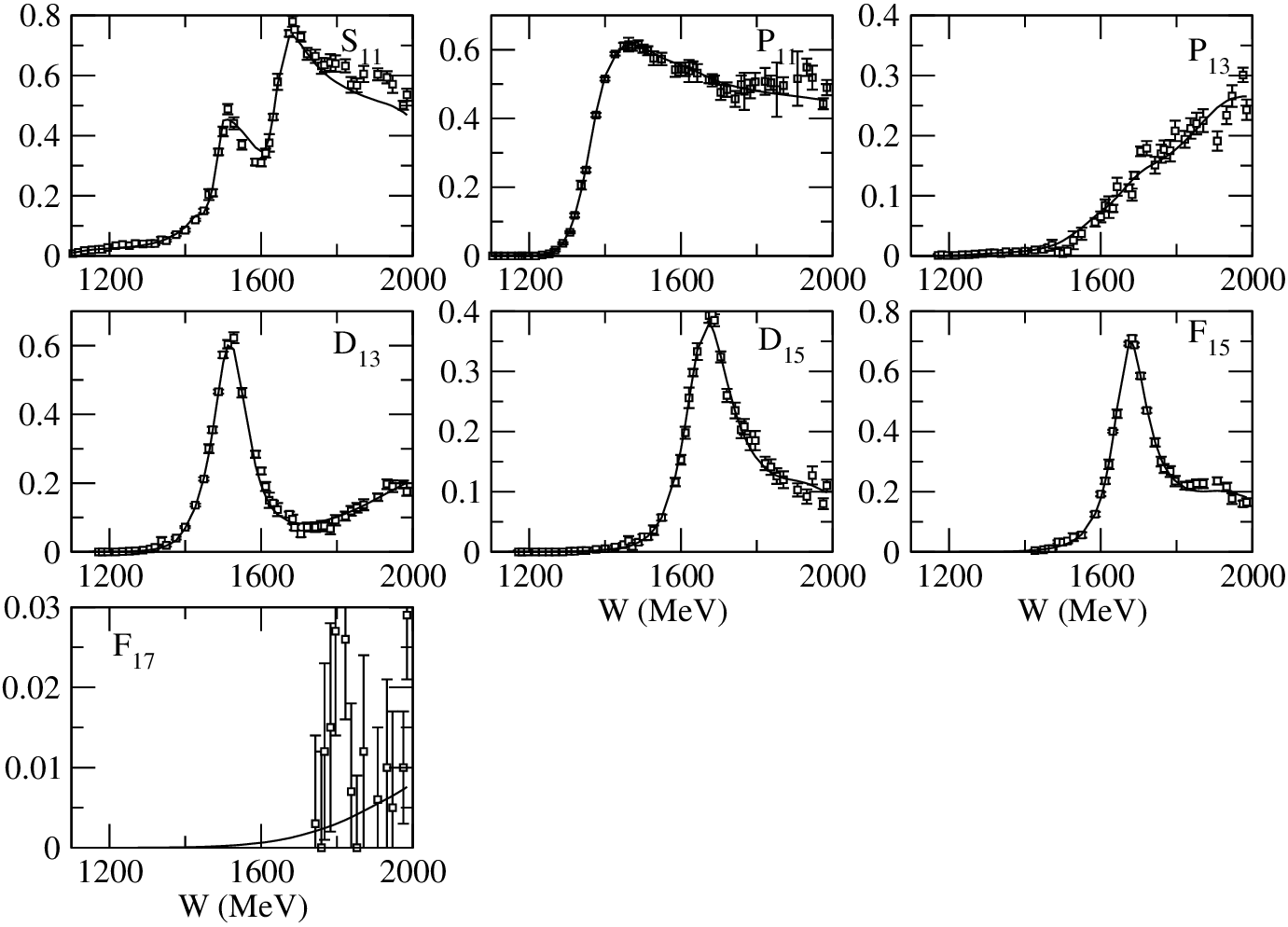, width=110mm}}
\end{center}
\caption{Imaginary parts of the calculated $\pi N$ partial wave amplitudes 
(Eq.~(\ref{eq:tmbmb})) of isospin $T=1/2$  are compared with
the energy independent solutions of Ref.~\cite{said}.  \label{figimagiso1}}
\end{figure}

\begin{figure}[tbp]
\begin{center}
\mbox{\epsfig{file=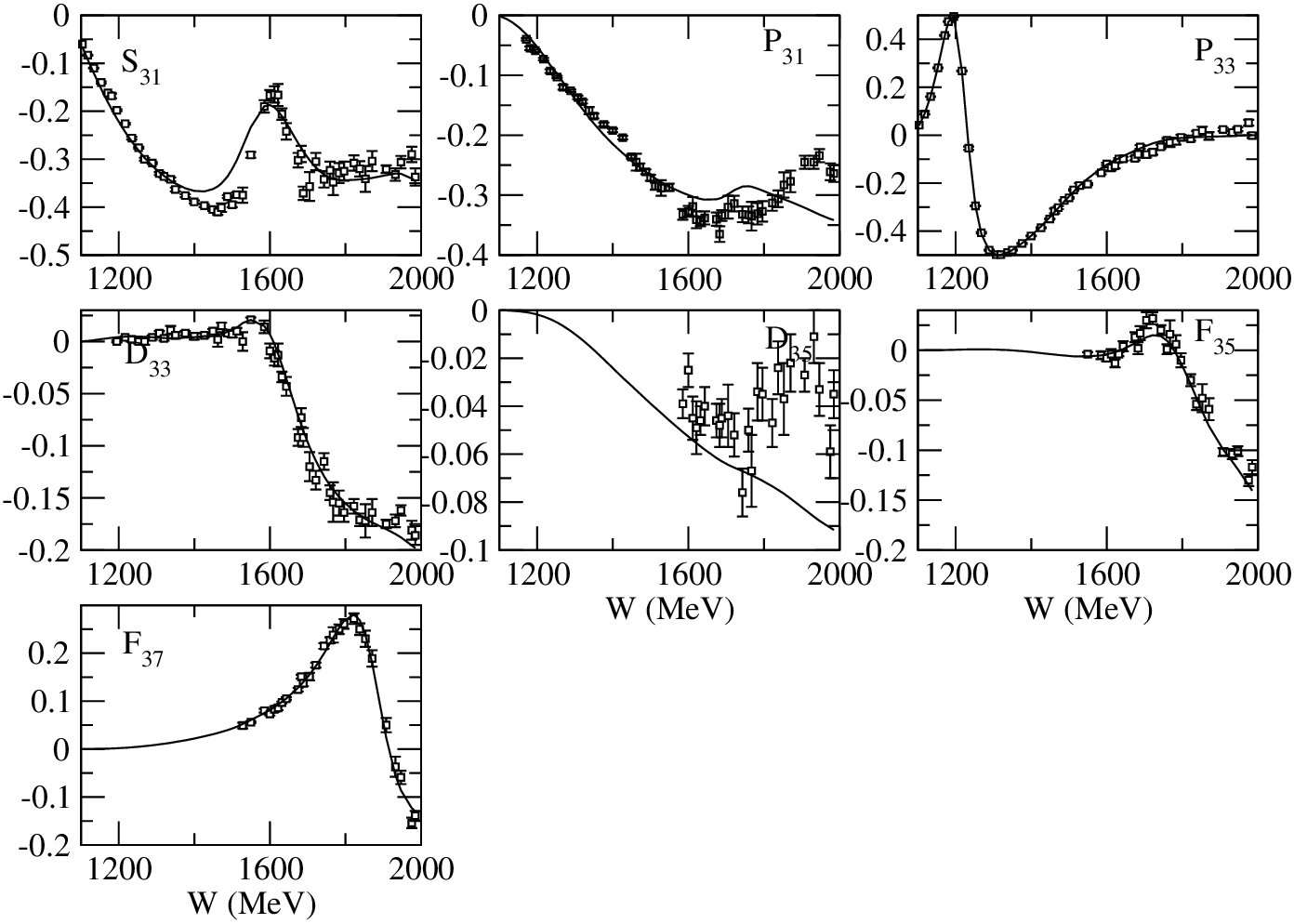, width=110mm}}
\end{center}
\caption{Real parts of the calculated $\pi N$ partial wave amplitudes
(Eq.~(\ref{eq:tmbmb})) of isospin $T=3/2$  are compared with
the energy independent solutions of Ref.~\cite{said}.  \label{figrealiso3}}
\end{figure}

\begin{figure}[tbp]
\vspace{20pt}
\begin{center}
\mbox{\epsfig{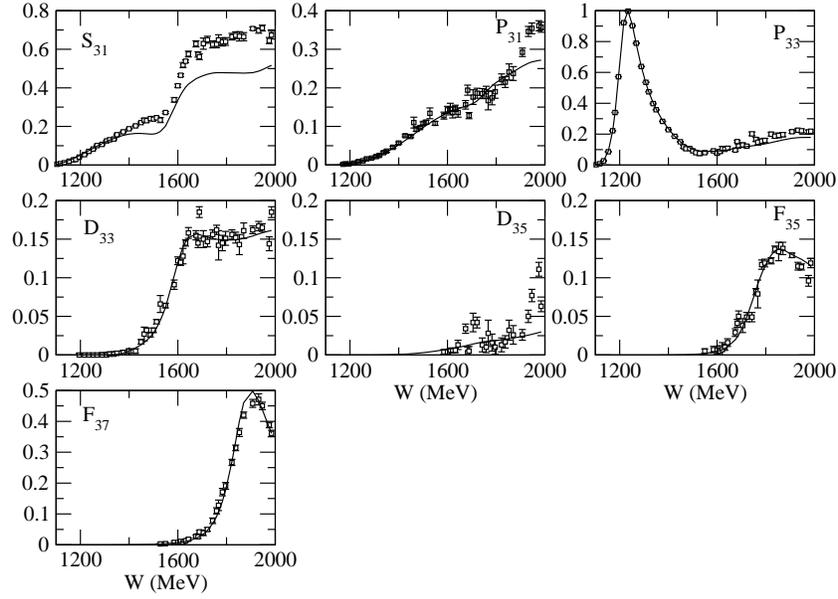}}
\end{center}
\caption{ Imaginary parts of the calculated $\pi N$ partial wave amplitudes
(Eq.~(\ref{eq:tmbmb})) of isospin $T=3/2$  are compared with
the energy independent solutions of Ref.~\cite{said}.
\label{figimagiso3}}
\end{figure}

Our fits to the empirical amplitudes of SAID~\cite{said} are given in
Figs.~\ref{figrealiso1}-\ref{figimagiso1} and Figs.~\ref{figrealiso3}-\ref{figimagiso3} 
for the $T=1/2$ and $T=3/2$ partial waves, respectively.
The resulting parameters are presented in Appendix I. The parameters associated 
with the non-resonant interactions, $v_{MB,M'B'}$ with $MB, M'B' = \pi N, \eta N$, 
$\pi\Delta, \rho N, \sigma N$, are given in Table~\ref{tab:nrcoup} for the coupling 
constants of the starting Lagrangian Eqs.(\ref{eq:L-pinn})-(\ref{eq:L-snn})
and Table~\ref{tab:nrcut} for the cutoffs of the form factors defined by
Eq.~(\ref{eq:ff}). The resulting bare $N^*$ parameters are listed in
Tables~\ref{tab:barem}-\ref{tab:RCU}

\begin{figure}[tbp]
\vspace{20pt}
\begin{center}
\mbox{\epsfig{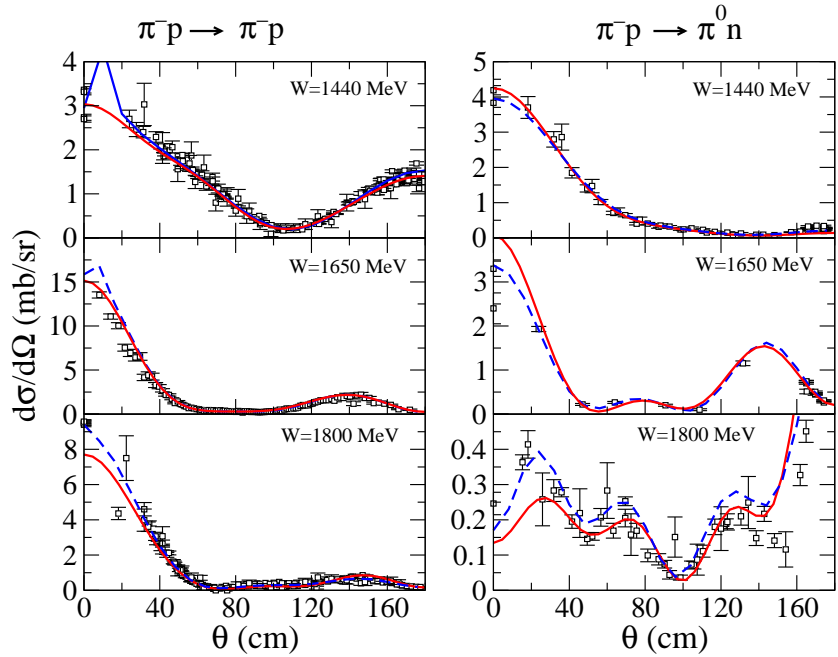}}
\end{center}
\caption{ Differential cross section for several
different center of mass energies. Solid red curve corresponds to our 
model while blue dashed lines correspond to the SP06 solution of 
SAID~\cite{said}. All data have been obtained through the SAID online
applications.~Ref.~\cite{said}.
\label{f11}}
\end{figure}

\begin{figure}[tbp]
\vspace{20pt}
\begin{center}
\mbox{\epsfig{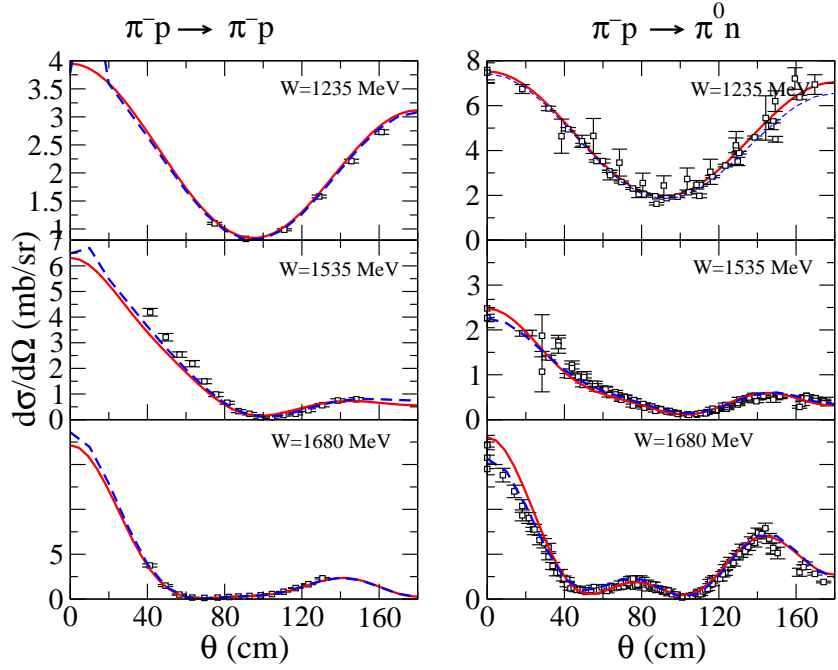}}
\end{center}
\caption{ Differential cross section for several
different center of mass energies. Similar description 
as Fig.~\ref{f11}. All
data have been obtained through the SAID online
applications.~Ref.~\cite{said}.
\label{f12}}
\end{figure}

\begin{figure}[tbp]
\vspace{20pt}
\begin{center}
\mbox{\epsfig{file=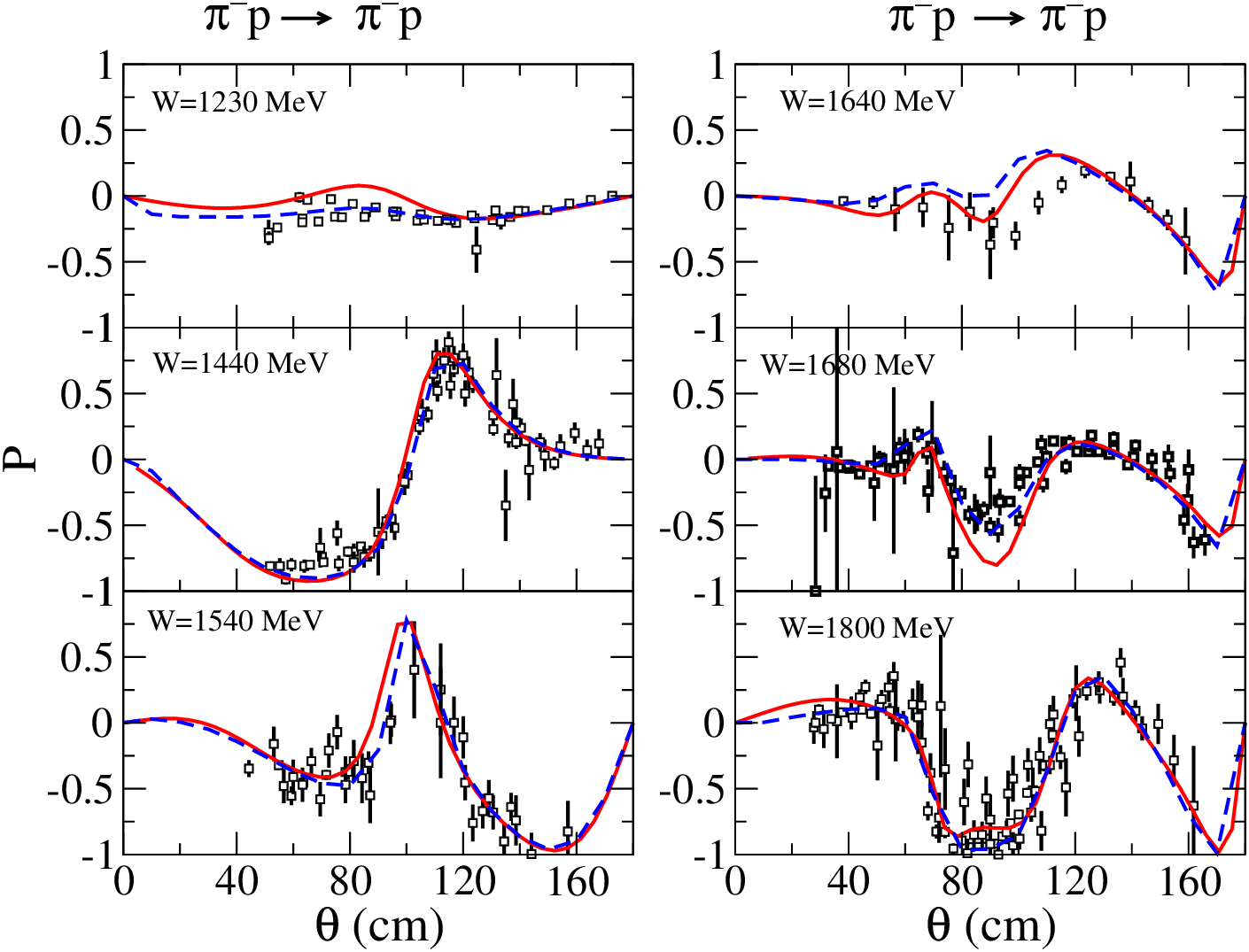, width=110mm}}
\end{center}
\caption{ Target polarization asymmetry, $P$, for several
different center of mass energies. Similar description 
as Fig.~\ref{f11}. All
data have been obtained through the SAID online
applications.~Ref.~\cite{said}.
\label{f13}}
\end{figure}

\begin{figure}[tbp]
\vspace{20pt}
\begin{center}
\mbox{\epsfig{file=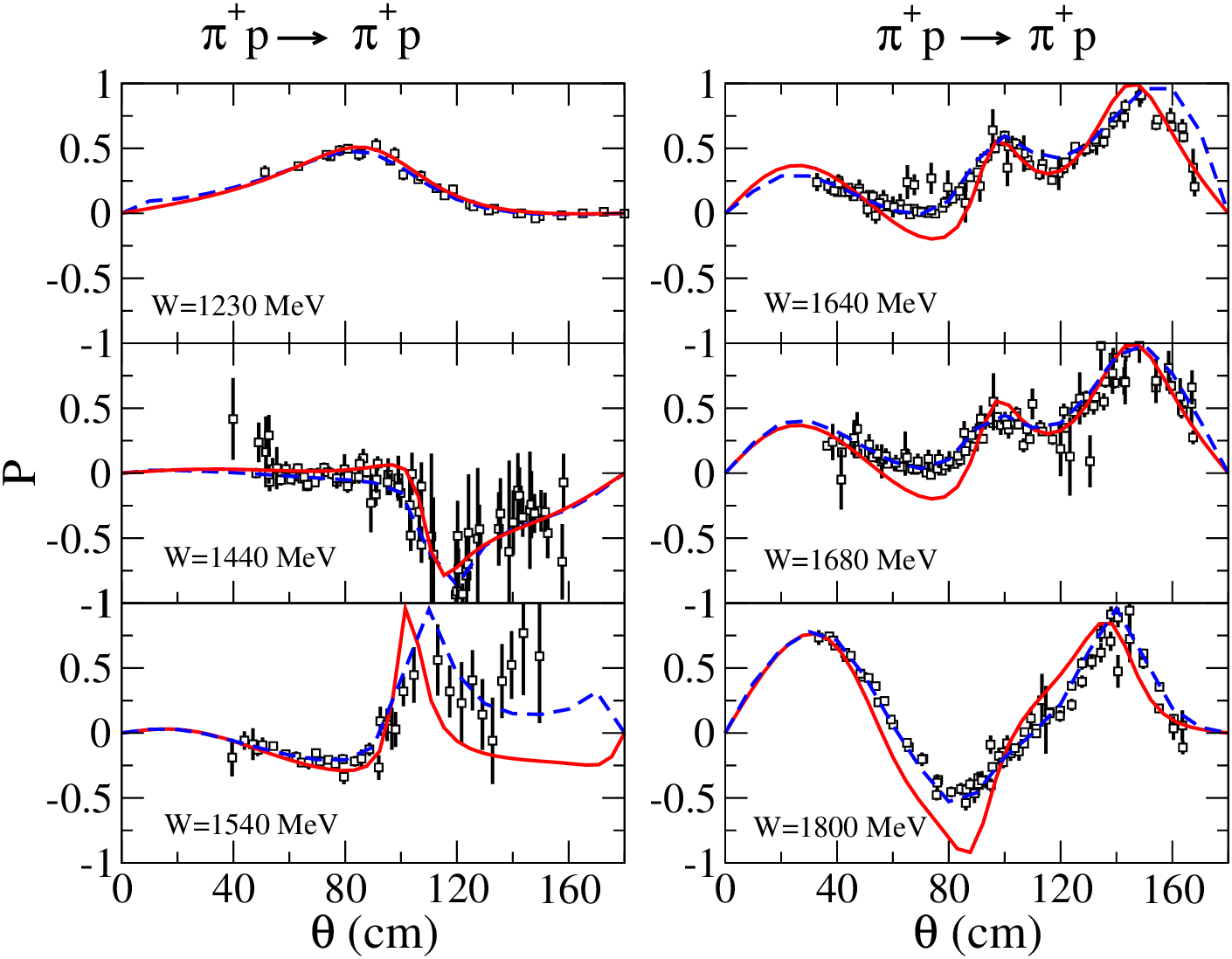, width=110mm}}
\end{center}
\caption{ Target polarization asymmetry, $P$, for several
different center of mass energies. Description 
as in Fig.~\ref{f11}. All
data have been obtained through the SAID online
applications.~Ref.~\cite{said}.
\label{f14}}
\end{figure}

\begin{figure}[tbp]
\vspace{20pt}
\begin{center}
\mbox{\epsfig{file=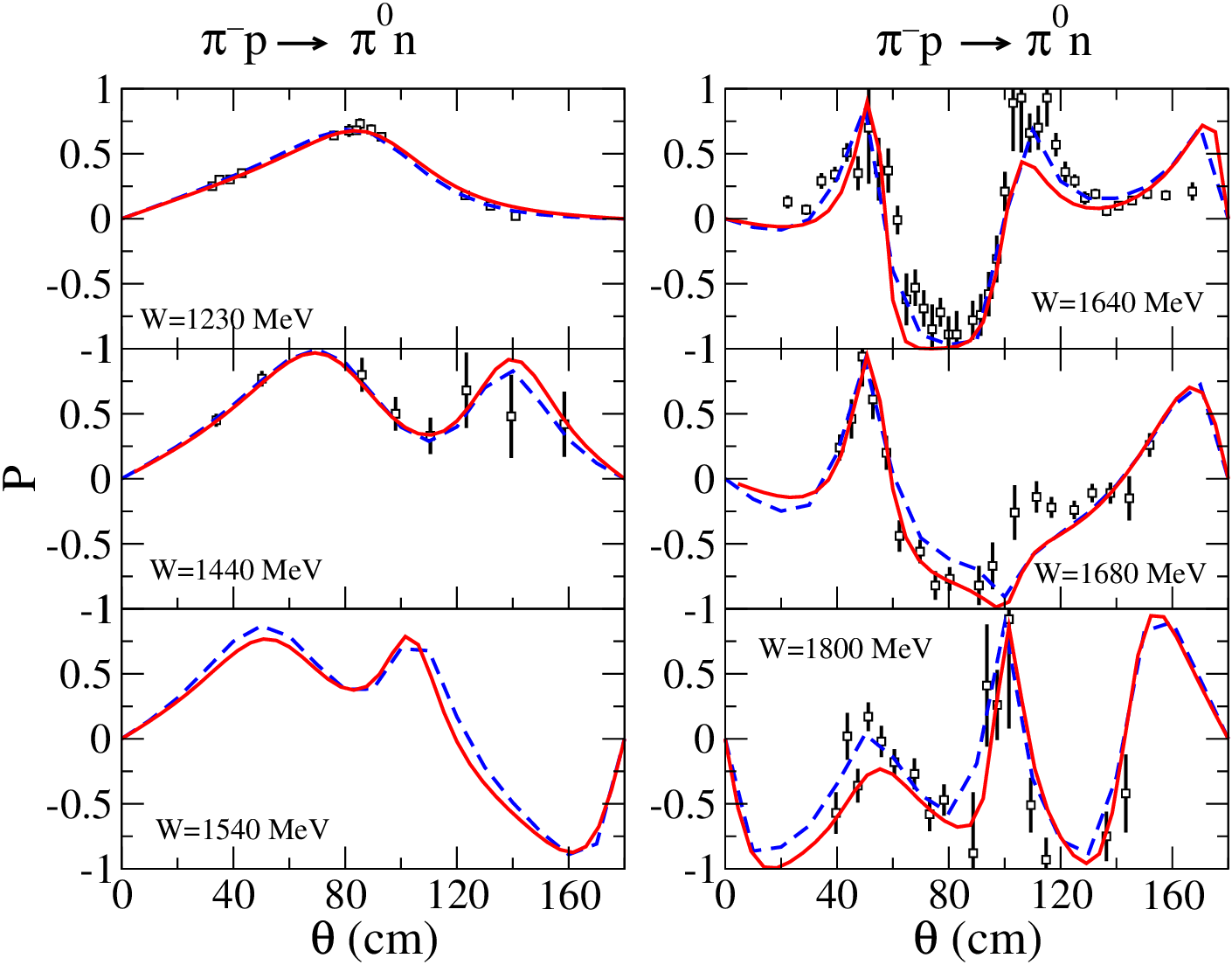, width=110mm}}
\end{center}
\caption{ Target polarization asymmetry, $P$, for several
different center of mass energies. Description 
as in Fig.~\ref{f11}. All
data have been obtained through the SAID online
applications.~Ref.~\cite{said}.
\label{f15}}
\end{figure}

From Figs.~\ref{figrealiso1}-\ref{figimagiso3}, one can see that the
empirical $\pi N$ amplitudes  can 
be fitted very well. The most significant discrepancies are
in the imaginary part of $S_{31}$ in 
Fig.\ref{figimagiso3}.
The agreement is also poor for the
$F_{17}$  in Fig.\ref{figrealiso1}-\ref{figimagiso1} and $D_{35}$ in
Figs.\ref{figrealiso3}-\ref{figimagiso3}, but there are 
rather large errors in the data. Our parameters are therefore
checked by directly comparing our predictions with the data of
 differential cross sections $d\sigma/d\Omega$ and target
polarization asymmetry $P$ of elastic $\pi^{\pm}p \rightarrow \pi^{\pm}p$
and charge-exchange $\pi^- p \rightarrow \pi^0 n$ processes.
Our results (solid red curves) are shown in Figs.\ref{f11}-\ref{f15}. 
Clearly, our model is rather consistent with the available data, and are
close to the results (dashed blue curves) calculated from the SAID's amplitudes.
Thus our model is justified despite the
differences with the SAID's amplitudes
seen in Fig.\ref{figrealiso1}-\ref{figimagiso3}.

It will be important to further refine our parameters by
fitting the data of other $\pi N$ scattering
observables, such as the recoil polarization and double polarization.
Hopefully, such  data  can be obtained from the
new hadron facilities at JPARC in Japan.


Our model is further checked by examining 
 our predictions of the total cross sections $\sigma^{tot}$ which
can be  calculated from the forward elastic scattering amplitudes by using 
the optical theorem. The total elastic scattering cross 
sections $\sigma^{el}$ 
can be calculated from the predicted partial wave amplitudes. With the 
normalization $<\vec{k}|\vec{k'}> =\delta (\vec{k}-\vec{k'})$ used 
in Ref.~\cite{msl}, we have
\begin{eqnarray}
\sigma^{el}(W) = \sum_{T=1/2,3/2} 
\sigma^{el}_T(W) 
\label{eq:crst-el1}
\end{eqnarray}
with
\begin{eqnarray}
\sigma^{el}_T(W) =\frac{(4\pi)^2}{k^2}\rho_{\pi N}(W) \sum_{JLS}
\frac{(2J+1)}{2} |T^{\, TJ}_{\pi N(LS), \pi N(LS)}(k,k,W)|^2 \,,
\label{eq:crst-el2}
\end{eqnarray}
where $\rho_{\pi N}(W)=\pi k E_\pi(k)E_N(k)/W$ with $k$ determined by 
$W=E_\pi(k) + E_N(k)$ and
the amplitude
$T^{TJ}_{L'S'(\pi N),LS(\pi N)}(k,k;W)$ is the partial-wave solution of 
Eq.~(\ref{eq:tmbmb}). Similarly, the total $\pi N \rightarrow \eta N$ cross sections
can be calculated from 
\begin{eqnarray}
\sigma^{tot}_{\pi N\rightarrow \eta N} = \frac{(4\pi)^2}{k^2}
\rho^{1/2}_{\pi N}(W) \rho^{1/2}_{\eta N}(W)\sum_{JLS} 
\frac{(2J+1)}{2} |T^{\, T=1/2, J}_{\eta N(LS), \pi N(LS)}(k',k,W)|^2 \,
\end{eqnarray}
where $\rho_{\eta N}(W)=\pi k' E_\eta(k')E_N(k')/W$ with $k'$ determined by
$W=E_\eta(k') + E_N(k')$.
We can also calculate the contribution from each of the
unstable channels, $\pi\Delta$, $\rho N$, and $\sigma N$, to the
total $\pi N \rightarrow \pi\pi N$ cross sections. For example, we have
for the $\pi N \rightarrow \pi \Delta \rightarrow \pi\pi N$ contribution in
the center of mass frame  
\begin{eqnarray}
\sigma^{rec}_{\pi \Delta}(W)= \int_{m_N+m_\pi}^{W-m_{\pi}}dM_{\pi N}
\frac{M_{\pi N}}{E_{\Delta}(k)}\frac{\Gamma_{\pi\Delta}(k,E)/(2\pi)}
{|W-E_\pi (k)-E_\Delta(k)-\Sigma_{\pi \Delta}(k,E)|^2}
\sigma_{\pi N \rightarrow \pi \Delta}(k,W) \label{eq:dcsine1}
\end{eqnarray}
where $k$ is defined by
$M_{\pi N}=E_\pi(k)+E_N(k)$,  
$E_{\pi N}(k)=[M^2_{\pi N} + k^2]^{1/2}$, 
$\Sigma_{\pi \Delta}(k,E)$ is defined in Eq.(\ref{eq:sig-pid}), 
$\Gamma_{\pi\Delta}(k,E)=-2 Im(\Sigma_{\pi \Delta}(k,E))$, and 
\begin{eqnarray}
\sigma_{\pi N \rightarrow \pi \Delta}(k,W)
&=&4\pi \rho_{\pi N}(k_0)\rho_{\pi\Delta}(k)\sum_{L'S',LS,J}
\frac{2J+1}{(2S_N+1)(2S_\pi +1)}
|T^J_{\pi\Delta(L'S'),\pi N (LS)}(k,k_0;W)|^2 \nonumber \\
& & \label{eq:dcsine2}
\end{eqnarray}
where $k_0$ is defined by $W=E_\pi(k_0)+E_N(k_0)$ and
 $\rho_{ab}(k)=\pi kE_a(k)E_b(k)/W$. The amplitude 
$T^J_{L'S'(\pi\Delta),LS(\pi N)}(k,k_0;W)$ is the partial-wave solution
of Eq.(\ref{eq:tmbmb}). The corresponding expressions for the 
unstable channels $\rho N$ and $\sigma N$ can be obtained 
from Eqs.~(\ref{eq:dcsine1})-(\ref{eq:dcsine2}) by changing 
the channel labels.

\begin{figure}[tbhp]
\vspace{20pt}
\begin{center}
\mbox{\epsfig{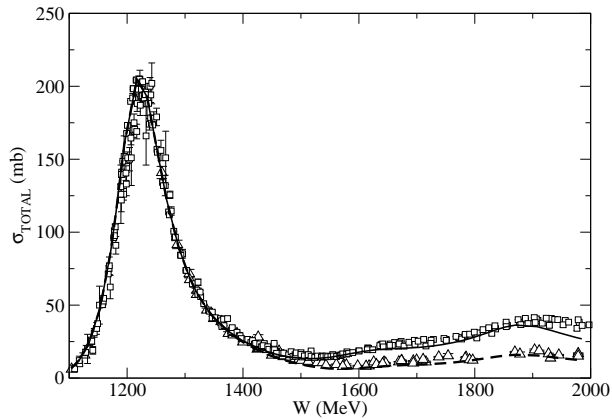}}
\end{center}
\caption{
The predicted total cross sections of the $\pi^+ p \to X $ (solid curve)and
$\pi^+ p \to \pi^+ p$ (dashed curve) reactions are compared with the data.
Squares and triangles are the corresponding data
from Ref.~\cite{PDG}.\label{pip}}
\end{figure}

\begin{figure}[th]
\vspace{20pt}
\begin{center}
\mbox{\epsfig{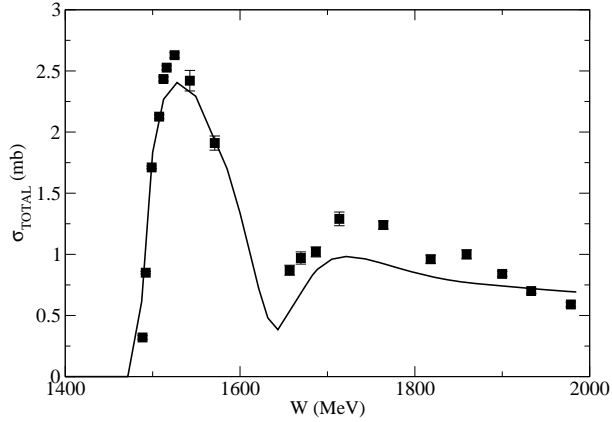}}
\end{center}
\caption{The predicted total cross sections of $\pi p \rightarrow \eta p$
reaction are compared with the data~\cite{eta-1,eta-2}.}
\label{pin-etan}
\end{figure}

The predicted $\sigma^{tot}$ (solid curves) along with the resulting
total elastic scattering cross sections $\sigma^{el}$ compared with 
the data of $\pi^+ p$ reaction are shown in Fig.~\ref{pip}.
Clearly, the model can
account for the data very well within the
experimental errors. Here
only the $T=3/2$ partial waves are relevant.
Equally good agreement with the data for $\pi^- p$ reaction are shown in 
the left side of Fig.~\ref{pim}.
In the right side, we show how the contributions from
each channel add up to get the total
cross sections. The comparison of the contribution from $\eta N$
channel with the data is shown in Fig.~\ref{pin-etan}.
It is possible to improve the fit to this data by adjusting
$N^* \rightarrow \eta N$ parameters. But this can be done correctly
only when the differential cross section data of $\pi N\rightarrow \eta N$
are included in the fit. This is beyond the scope of this work and will
be pursed in our second-stage calculations.

The contributions from $\pi \Delta$, $\rho N$ and $\sigma N$
intermediate states to the $\pi^- p \rightarrow \pi\pi N$ total cross sections
calculated from our model can be seen 
in the right side of Fig.~\ref{pim}. These predictions
remain to be verified by the future experiments. The existing 
$\pi N \rightarrow \pi\pi N$ data are not sufficient for extracting 
$model$ $independently$ the contributions from each unstable channel.

The results shown in 
Figs.~\ref{pip}-\ref{pin-etan} indicate that our parameters are consistent 
with the total cross section data.

\begin{figure}[tbhp]
\vspace{25pt}
\begin{center}
\mbox{\epsfig{file=fig14.eps, width=120mm}}
\end{center}
\caption{Left: The predicted total cross sections of the $\pi^- p\to X$ (solid curve)
and $\pi^-p \to \pi^- p + \pi^0 n$ (dashed curve) reactions
are compared with the data. 
Open squares are the  data on $\pi^-p\to X$ 
from Ref.~\cite{PDG}, open triangles are obtained by adding the 
 $\pi^-p \to \pi^-p$ and $\pi^-p\to \pi^0n$ data obtained 
from Ref.~\cite{PDG} and SAID database~\cite{saiddb} respectively.
Right: Show how the predicted contributions from each channel are added up
to the predicted total cross sections of the $\pi^- p\to X$.
\label{pim}}
\end{figure}

We now discuss the parameters presented in Appendix A.
It is rather difficult to compare
the resulting non-resonant coupling constants
listed in Table~\ref{tab:nrcoup}
with the values from other works,
since the coupling strengths are also determined by the cutoff
parameters listed in Table~\ref{tab:nrcut}.
Perhaps it is possible to narrow their
differences by using a different parameterization of the form factors.
However, the fit is a rather time consuming process and
hence no attempt is made in this work to try other forms of
form factors.

In Table~\ref{tab:barem}, we see that all of the bare masses are higher than
the PDG's resonance positions. This can be understood from the expression
Eq.~(\ref{eq:bw-er}) for the partial waves with only one $N^*$ since
one finds in general that $Re[\bar{\Sigma}(E)] < 0$. For the
$S_{11}$, $P_{11}$, $P_{33}$ and $D_{13}$ partial waves, two bare $N^*$
states are mixed by their interactions, as can be seen in Eq.~(\ref{eq:nstar-sigma}).
Thus the relation between their bare masses and the resonance positions
identified by PDG is much more complex.

As we mentioned above, the fit to $\pi N$ elastic scattering is can not
determine well the bare $N^* \rightarrow \pi \Delta, \rho N, \sigma N$
parameters. Thus the results for these unstable particle channels listed
in Tables~\ref{tab:nrcoup}-\ref{tab:RCU} must be refined by fitting
the $\pi N \rightarrow \pi\pi N$ data.

\section{Summary and Future Developments}

Within the formulation developed in Ref.~\cite{msl}, we have constructed 
a dynamical coupled-channel model of $\pi N$ scattering by fitting the 
$\pi N$ scattering data.
The parameters of the model are first determined by fitting as much as
 possible the empirical $\pi N$ elastic scattering amplitudes
of SAID up to 2 GeV. We then refine and confirm the resulting parameters by
directly comparing the predicted differential cross section and
target polarization asymmetry
with the original data of the elastic $\pi^{\pm} p \rightarrow \pi^{\pm} p$
and charge-exchange $\pi^- p \rightarrow \pi^0 n$ processes.
The predicted total cross sections of $\pi N$ reactions and
are also in good agreement with the data.
The model thus can be 
used as a starting point for analyzing the very extensive data of 
electromagnetic $\pi$ production reactions. 

The predicted total cross sections of $\pi N\rightarrow \eta N$ reactions
are also in fair agreement with the data.
However, the parameters associated with the $\eta N$ channel 
need to be refined to also fit the differential cross section data 
of $\pi N \rightarrow \eta N$ before the model can be used to analyze 
the data of electromagnetic $\eta$ production reactions. 

The main shortcoming of this work is that the $\pi\pi N$ interaction term 
$h_{\pi \pi N}$ of Eq.(\ref{eq:hpipin}) is not included 
in the calculations. As derived in Ref.~\cite{msl}, the effects due to this
interaction can be included by adding a term $Z^{(I)}_{MB,M'B'}(E)$,
which contains the $\pi\pi N \rightarrow \pi\pi N$ scattering amplitude, 
to the driving term $V_{MB,M'B'}(E)$ of Eq.(\ref{eq:nr-tmbmb}). 
Our effort in this direction is in progress along with the
development of a more complete determination of the parameters of the model
by fitting both the data of $\pi N$ elastic 
scattering and $\pi N \rightarrow \pi\pi N$ reactions.
This is also essential to pin down the parameters of the interactions
associated with the $\pi\Delta$, $\rho N$ and $\sigma N$ states.
 Only when this 
second-stage is completed, we then can perform dynamical 
coupled-channel analysis of the very extensive and complex data of 
photo- and electro-production of two pions. This is an essential step 
to probe the $W > $ about 1.7 GeV resonance region where the information 
on $N^*$ is very limited and uncertain.
   
Finally, a necessary next step is to extract the resonance 
poles and the associated residues from the 
predicted $\pi N$ amplitudes. This is being pursued and will be 
published elsewhere~\cite{suzukikun}.

\begin{acknowledgments}
We would like to thank M. Paris for his assistance in using the parallel 
processors at NERSC and A. Parre\~no for her help and encouragement to use
the BSC. This work is supported by the U.S. Department of Energy, 
Office of Nuclear Physics Division, under contract No. DE-AC02-06CH11357, 
and Contract No. DE-AC05-060R23177 
under which Jefferson Science Associates operates Jefferson Lab,
and by the Japan Society for the Promotion of Science,
Grant-in-Aid for Scientific Research(c) 15540275. This work is also
partially supported by Grant No. FIS2005-03142 from MEC (Spain) 
and FEDER and European Hadron Physics Project RII3-CT-2004-506078. 
The computations were performed at NERSC (LBNL) and Barcelona 
Supercomputing Center (BSC/CNS) (Spain). The authors thankfully acknowledges 
the computer resources, technical expertise and assistance provided by the 
Barcelona Supercomputing Center - Centro Nacional de Supercomputacion (Spain).
\end{acknowledgments}


\appendix

\section{Parameters from the fits}

\begin{table}[bp]
\begin{ruledtabular}
\begin{tabular}{c|l|l|l  }
Parameter              &               & SL Model     \\
\hline
$f^2_{\pi NN}/(4\pi)$ & 0.08          &    0.08 \\
$m_{\sigma}$ (MeV)     &  500.1         &$-$\\
$f_{\pi N\Delta}$      &  2.2061         & 2.0490 \\
$f_{\eta NN}$           &  3.8892        &$-$\\
$g_{\rho NN}$           &  8.7214        & 6.1994 \\
$\kappa_{\rho}$             &  2.654           &1.8250\\
$g_{\omega NN}$         &  8.0997         & 10.5 \\
$\kappa_{\omega}$           &  1.0200         & 0.0 \\
$g_{\sigma NN}$         &  6.8147         &$-$\\
$g_{\rho \pi\pi}$        &  4.            & 6.1994 \\
$f_{\pi\Delta\Delta}$  &  1.0000         &$-$ \\
$f_{\rho N\Delta}$     &  7.516          &$-$ \\
$g_{\sigma \pi\pi }$    &  2.353           &$-$\\
$g_{\omega \pi \rho}$  &  6.955          &$-$ \\
$g_{\rho \Delta\Delta}$&  3.3016         &$-$ \\
$k_{\rho\Delta\Delta}$ & 2.0000          &$-$  \\
\end{tabular}
\caption{The parameters associated with the Lagrangians Eqs.(22)-(35).
The results are from fitting 
the empirical $\pi N$ partial-wave amplitudes~\cite{said} of a given total
isospin $T= 1/2$ or $3/2$.
The parameters from the SL model of Ref.~\cite{sl-1} are
also listed.
\label{tab:nrcoup}}
\end{ruledtabular}
\end{table}

\begin{table}[tbp]
\begin{ruledtabular}
\begin{tabular}{c|l|l|l  }
Parameter                       &         (MeV)                 & SL model
(MeV)   \\
\hline
$\Lambda_{\pi NN}$               &  809.05     & 642.18 \\
$\Lambda_{\pi N\Delta}$          &  829.17     &648.18 \\
$\Lambda_{\rho NN}$              &  1086.7     &1229.1 \\
$\Lambda_{\rho\pi\pi}$          &  1093.2    &1229.1\\
$\Lambda_{\omega NN}$            &  1523.18    &$-$ \\
$\Lambda_{\eta NN}$              &  623.56     &$-$ \\
$\Lambda_{\sigma NN}$            &  781.16     &$-$ \\
$\Lambda_{\rho N\Delta}$         &  1200.0     &$-$ \\
$\Lambda_{\pi \Delta\Delta}$     &  600.00     &$-$ \\
$\Lambda_{\sigma\pi\pi}$        &  1200.0      &$-$ \\
$\Lambda_{\omega\pi\rho}$       &  600.00      &$-$ \\
$\Lambda_{\rho \Delta\Delta}$    &  600.00     &$-$ \\
\end{tabular}
\caption{Cut-offs of the form factors, Eq.~(\ref{eq:ff}), of 
the non-resonant interaction $v_{MB,M'B'}$. 
The results are from fitting
the empirical $\pi N$ partial-wave amplitudes~\cite{said} of a given total
isospin $T= 1/2$ or $3/2$.
The parameters from the SL model of Ref.~\cite{sl-1} are
also listed.
\label{tab:nrcut}}
\end{ruledtabular}
\end{table}


\begin{table}[tbp]
\begin{ruledtabular}
\begin{tabular}{l|l|ll  }
 $L_{TJ}$& PDG's Mass( MeV)           &  $M_1$ (MeV) &  $M_2$ (MeV)  \\
\hline
  $S_{11}$& 1535; 1655  &    1800.     &   1880.      \\ 
  $S_{31}$& 1630 &   1850.      &       \\ 
  $P_{11}$& 1440; 1710 &    1763     &   2037      \\ 
  $P_{13}$& 1720 &    1711     &       \\ 
  $P_{31}$& 1910 &   1900.3     &               \\ 
  $P_{33}$& 1232; 1600 &     1391    &   1602.     \\ 
  $D_{13}$& 1520; 1700 &      1899.1   &    1988.     \\ 
  $D_{15}$& 1675 &    1898     &               \\ 
  $D_{33}$& 1700 &  1976       &               \\ 
  $D_{35}$& 1960 &   $-$     &               \\ 
  $F_{15}$& 1685 &    2187    &               \\ 
  $F_{35}$& 1890 & 2162        &               \\ 
  $F_{37}$& 1930 &  2137.8       &               \\ 
\end{tabular}
\caption{The masses of the nucleon excited states included in the fits.
(second and third 
columns). The first column contains the 
masses of  the nucleon
resonances given by PDG~\cite{PDG}.\label{tab:barem}}
\end{ruledtabular}
\end{table}

\begin{table}[tbp]
\begin{ruledtabular}
\begin{tabular}{c||l||l||l|l||l||l|l|l}
            &   $\pi N$   & $\eta N$  &  \multicolumn{2}{c||}{$\pi \Delta$}  &  $\sigma N$  & \multicolumn{3}{c}{$\rho N$} \\
\hline
  $S_{11}$  (1)&  7.0488      &  9.1000   &     $-$1.8526  &            &  $-$2.7945      &     2.0280 &   .02736 &             \\
  $S_{11}$  (2)& 9.8244       & .60000    &  .04470      &            &     1.1394      &  $-$9.5179 &$-$3.0144 &             \\
  $S_{31}$     & 5.275002     & $-$       &  $-$6.17463    &            &  $-$          &$-$4.2989 & 5.63817    &           \\
  $P_{11}$  (1)&   3.91172    &   2.62103 &  $-$9.90545    &            & $-$7.1617       &$-$5.1570 &  3.45590&            \\
  $P_{11}$  (2)&    9.9978    &      3.6611 &  $-$6.9517     &            & 8.62949       &$-$2.9550 &$-$0.9448 &          \\
  $P_{13}$     &   3.2702     & $-$.99924   &  $-$9.9888     &   $-$5.0384  &   1.0147      &$-$.00343 & 1.9999   & $-$.08142     \\
  $P_{31}$     &  6.80277     &$-$        &  2.11764     &            &   $-$         & 9.91459    &0.15340   &             \\
  $P_{33}$ (1) &  1.31883     & $-$       &  2.03713     & 9.53769    & $-$           &$-$.3175  &1.0358    &  0.76619    \\
  $P_{33}$ (2) &  1.3125      &  $-$      &  1.0783        & 1.52438      &$-$            & 2.0118   &$-$1.2490 &  0.37930      \\ 
  $D_{13}$ (1) &  .44527      & $-$.0174  &  $-$1.9505     &  .97755      & $-$.481855    & 1.1325   & $-$.31396 &.17900      \\ 
  $D_{13}$ (2) &  .46477      & .35700    &  9.9191      &  3.8752    &  $-$5.4994      &  .28916  & 9.6284 & $-$.14089     \\
  $D_{15}$     &  .31191      & $-$.09594 &   4.7920     &  .01988    &    $-$.45517    & $-$.17888  &  1.248& $-$.10105     \\
  
  $D_{33}$     &  .9446       &   $-$     &  3.9993       &  3.9965     &    $-$        & .16237     & 3.948  & $-$.85580     \\  
  $F_{15}$     & .06223       & 0.0000    & 1.0395      &   .00454      &   1.5269      & $-$1.0353  &1.6065  & $-$.0258      \\
  $F_{35}$     &  .173934     &   $-$     & $-$2.96090    &  $-$1.09339 &    $-$        &$-$.07581 & 8.0339 & $-$.06114     \\
  $F_{37}$     &  0.25378     &   $-$     & $-$0.3156     & $-$0.0226   &    $-$        & .100     & .100   & .100        \\
\end{tabular}
\caption{The coupling constants $C_{N^*,JTLS;MB}$ of Eq.~(\ref{eq:gmb})
with $MB = \pi N, \eta N, \pi\Delta,\sigma N, \rho N$ for each of 
the resonances. 
When there are more than one value for $\pi\Delta$ and $\rho N$
channels, they correspond to 
the possible quantum numbers $(LS)$ listed in
Table 2.
\label{tab:RC}}
\end{ruledtabular}
\end{table}


\begin{table}
\begin{ruledtabular}
\begin{tabular}{c||l||l||l|l||l||l|l|l}
            &   $\pi N$   & $\eta N$  &  \multicolumn{2}{c||}{$\pi \Delta$}  &  $\sigma N$  & \multicolumn{3}{c}{$\rho N$} \\                                                                                    
\hline
  $S_{11}$  (1)&   1676.4     &  598.97  &      554.04  &            &  801.03       &  1999.8  & 1893.6  &            \\
  $S_{11}$  (2)&  533.48      &  500.02    &  1999.1      &            & 1849.5     &  796.83  & 500.00 &             \\
  $S_{31}$     &  2000.00     & $-$       &   500.00     &            &  $-$          &  500.031 & 500.00 &             \\
  $P_{11}$  (1)&    1203.62   &    1654.85&    729.0     &            &  1793.0       &  621.998 & 1698.90&             \\
  $P_{11}$  (2)&     646.86   &    897.84 &   501.26     &            &  1161.20      &  500.06  &  922.280&            \\
  $P_{13}$     &    1374.0    &  500.23   &   500.00     &   500.770  &    640.50     &   500.00 &  500.10& 1645.2     \\
  $P_{31}$     &   828.765    &$-$        &  1999.9      &            &   $-$         &  1998.8  & 2000.6 &             \\
  $P_{33}$ (1) &  880.715     & $-$       &  507.29      &  501.73    & $-$           & 606.78   & 1043.4 &  528.37     \\
  $P_{33}$ (2) &  746.205     &  $-$      &  846.37      &  780.96    &$-$            &  584.98  & 500.240& 1369.7      \\ 
  $D_{13}$ (1) &  1658.       &   1918.2  &   976.36     &   1034.5   &  1315.8       &  599.79  &1615.1 & 1499.50     \\ 
  $D_{13}$ (2) &   1094.0     &  678.41   &   1960.0     &  660.02    &   1317.0      &  550.14  &  597.57&  1408.7     \\
  $D_{15}$     &   1584.7     &  1554.0   &   500.77     &   820.17   &    507.07     & 735.40   &  749.41  & 937.53   \\
  $D_{33}$     & 806.005      &   $-$     &  1359.38    &  608.090    &  $-$          & 1514.98  &1998.99 & 956.61      \\ 
  $F_{15}$     &  1641.6      & 655.87    & 1899.5      &   522.68   &   500.93      &   500.76 &  500.0&   1060.9    \\
  $F_{35}$     &   1035.28    &    $-$    &  1227.999   &    586.79   &   $-$         & 1514.3   & 593.84 & 1506.0      \\
  $F_{37}$     &   1049.04    &    $-$    &  1180.2     &    1031.81  &   $-$         & 600.02   & 600.00 & 600.02      \\
\end{tabular}
\caption{The range parameter $\Lambda_{N^*,JTLS;MB}$ (in unit of (MeV/c)) 
of Eq.~(\ref{eq:gmb})
with $MB = \pi N, \eta N, \pi\Delta,\sigma N, \rho N$ for each of
the resonances.
When there are more than one value for $\pi\Delta$ and $\rho N$ channels, 
they correspond to the possible quantum 
numbers ($LS$) listed in
Table 2.
\label{tab:RCU}
}
\end{ruledtabular}
\end{table}


\begin{thebibliography}{99}
\bibitem{propag}
The  Lorentz boost factors for transforming the matrix elements of
$\Delta \rightarrow \pi N$ ($\rho, \sigma \rightarrow \pi\pi$) 
vertex interaction from the $\pi N$ ($\pi\pi$) to the
$\pi\pi N$ center of mass frames, which were not given in Ref.~\cite{msl}, 
are included here.
\end{thebibliography}
\end{document}